\documentclass[journal=jacsat,manuscript=article]{achemso}

\usepackage{chemformula} 
\usepackage[T1]{fontenc} 
\usepackage{multirow}
\SectionNumbersOn

\author{Setthanat Wijitpatima}
\affiliation[TU Berlin]
{Institute of Solid State Physics, Technische Universität Berlin, Hardenbergstraße 36, Berlin 10623, Germany}

\author{Normen Auler}
\affiliation[Paderborn]
{Department of Physics, Paderborn University, Warburger Str. 100, 33098 Paderborn, Germany}

\author{Priyabata Mudi}
\author{Timon Funk}
\author{Avijit Barua}
\affiliation[Berlin]
{Institute of Solid State Physics, Technische Universität Berlin, Hardenbergstraße 36, Berlin 10623, Germany}

\author{Binamra Shrestha}
\affiliation[Paderborn]
{Department of Physics, Paderborn University, Warburger Str. 100, 33098 Paderborn, Germany}

\author{Imad Limame}
\author{Sven Rodt}
\affiliation[TU Berlin]
{Institute of Solid State Physics, Technische Universität Berlin, Hardenbergstraße 36, Berlin 10623, Germany}

\author{Dirk Reuter}
\affiliation[Paderborn]
{Department of Physics, Paderborn University, Warburger Str. 100, 33098 Paderborn, Germany}

\author{Stephan Reitzenstein}
\email{stephan.reitzenstein@physik.tu-berlin.de}
\affiliation[TU Berlin]
{Institute of Solid State Physics, Technische Universität Berlin, Hardenbergstraße 36, Berlin 10623, Germany}

\title[Manuscript]
  {Bright electrically contacted circular Bragg grating resonators with deterministically integrated quantum dots}

\begin{document}

\begin{tocentry}
\includegraphics[width=10cm]{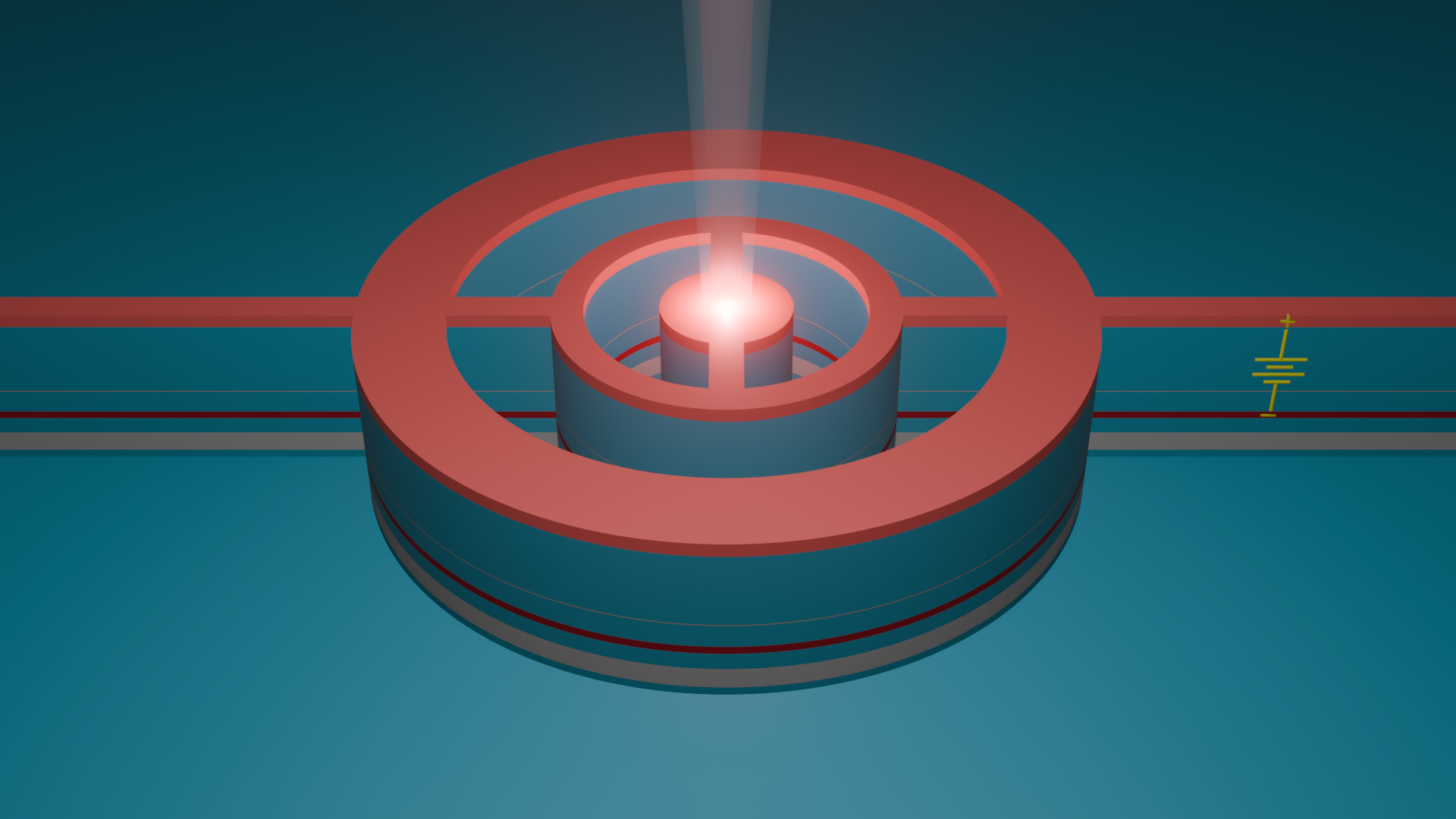}
\end{tocentry}


\begin{abstract}
Cavity-enhanced emission of electrically controlled semiconductor quantum dots is essential in developing bright quantum devices for real-world quantum photonic applications. Combining the circular Bragg grating (CBG) approach with a PIN-diode structure, we propose and implement an innovative concept for ridge-based electrically-contacted CBG resonators. Through fine-tuning of device parameters in numerical simulations and deterministic nanoprocessing, we produced electrically controlled single quantum dot CBG resonators with excellent electro-optical emission properties. These include multiple wavelength-tunable emission lines and a photon extraction efficiency (PEE) of up to (30.4±3.4)\%, where refined numerical optimization based on experimental findings suggests a substantial improvement, promising PEE >50\%. Additionally, the developed quantum light sources yield single-photon purity reaching (98.8±0.2)\% [post-selected: (99.5±0.3)\%] and a photon indistinguishability of (25.8±2.1)\% [post-selected: (92.8±4.8)\%]. Our results pave the way for high-performance quantum devices with combined cavity enhancement and deterministic charge-environment controls, advancing the development of photonic quantum information systems such as complex quantum repeater networks. 
\end{abstract}

\section{Introduction}
In the field of photonic quanatum information processing the concept of quantum repeaters \cite{briegel_quantum_1998} has emerged as a cornerstone of advanced quantum communication, serving as a key element to extend the communication range limited in simple point-to-point concepts by absorption to distances below about 100 km, and thereby enabling high bit-rate long-range transmission \cite{sangouard_long-distance_2007}. Implementing quantum repeater networks requires on-demand single-photon sources with high single-photon purity and indistinguishability, as well as wavelength tunability and electrically controllable spin-photon interfaces \cite{lu_push-button_2014, wang_-demand_2019}. These stringent criteria have driven extensive research efforts, encompassing the exploration of suitable material systems and tailoring practical device designs for real-world applications \cite{aharonovich_solid-state_2016}. In this context, semiconductor quantum dots (QDs) have been the most outstanding candidates for single-photon sources in quantum information technology scenarios \cite{heindel_quantum_2023}. Not only do they exhibit excellent quantum optical properties, but based on semiconductor materials, they also allow for device integration using advanced deterministic nanofabrication techniques, thus opening up possibilities to maximize their optical performance \cite{rodt_deterministically_2020}. 

For applications in photonic quantum information technology, external control over the electronic states of QDs is of great importance, for instance, to bring QDs in remote quantum light sources into spectral resonance which is needed to enable entanglement distribution via Bell-state measurements in quantum repeater networks \cite{patel_two-photon_2010,zhai_quantum_2022}. In this aspect, researchers have reported results on electrical charge control of QDs embedded within field-effect structures such as PIN diodes \cite{warburton_charged_1997,hermannstadter_polarization_2009,lobl_narrow_2017,schall_bright_2021}, and applied quantum-confined Stark effect \cite{miller_band-edge_1984} for spectral fine-tuning \cite{bennett_giant_2010,nowak_deterministic_2014,schnauber_spectral_2021}.

Parallel studies have focused on enhancing the photon extraction efficiency (PEE) of QD quantum light sources, for instance, via nanophotonic cavities. Among these cavities, circular Bragg grating (CBG) resonators have gained prominence due to their broadband emission enhancement in combination with pronounced light-matter interaction in the Purcell regime of cavity quantum electrodynamics \cite{davanco_circular_2011,liu_solid-state_2019,wang_-demand_2019}. A CBG resonator is typically fabricated by etching a series of circular trenches around a central disk with the embedded targeted QD, creating a high refractive index contrast to realize tight lateral confinement of the light field \cite{ates_bright_2012,yao_design_2018}. When a CBG resonator is paired with a back-side mirror, such as gold or a distributed Bragg reflector (DBR), redirecting the emitted photons upward, the vertical collection efficiency can be significantly boosted to PEE values exceeding 60\% in experiments \cite{liu_solid-state_2019,wang_-demand_2019}. However, due to the described geometry of a CBG resonator, a QD integrated inside such a structure is fundamentally electrically isolated. As a consequence, while the CBG-QD devices have performed well as on-demand devices under optical pumping \cite{liu_solid-state_2019,wang_-demand_2019}, the experimental demonstration of electrically-contacted CBG resonators has remained elusive \cite{rickert_high-performance_2023}. Notably, electrically-contacted QD molecule devices with a PIN-diode design and surface CBG resonators were demonstrated, achieving experimental PEEs up to 24\%, limited by only a partial utilization of the CBG concept \cite{schall_bright_2021}. 

To solve the issue of implementing high brightness CBG single-photon sources with electrical control of integrated QDs, recent design efforts have explored ridge-based CBG approaches \cite{ji_design_2021,barbiero_design_2022}. In these methods, instead of etching complete circular trenches around QDs as done on optically pumped structures, narrow ridges are unetched to retain the connection of doped layers in the central mesa to the area outside of the resonator where metallic pads can be flexibly prepared for external electrical control (see schematic drawings in Figure~\ref{fgr:device_design}). However, to date, only numerical designs and theoretical discussions of the optical performance of such modified CBGs have been reported \cite{buchinger_optical_2023,rickert_high-performance_2023,Shih2023}. 

Therefore, taking an important step forward, we present the first proof-of-concept experimental results on electrically-controlled QDs in CBG resonators. We first discuss aspects of designing an electrically-contacted CBG (eCBG) along with numerical simulations to maximize its PEE, then discuss the device fabrication and the optical characterization. The design utilizes a DBR as the back-side mirror to simplify the device fabrication process, instead of using a gold mirror, which requires a post-growth flip-chip process \cite{liu_solid-state_2019,wang_-demand_2019}. In this scenario, the heterostructure including DBR, QDs, and the PIN diode can be grown epitaxially in a single process using metal-organic chemical vapor deposition or molecular beam epitaxy. The deterministic nanofabrication process utilizes marker-based electron-beam lithography (EBL) in combination with low-temperature cathodoluminescence (CL) mapping to determine the positions and spectral features of suitable QDs before device integration. To investigate the electro-optical performance of the fabricated eCBG devices and assess the quantum optical properties of the deterministically integrated QDs, bias-voltage dependent micro-photoluminescence ($\mu$PL) measurements and time-resolved photon correlation measurements are performed. 

\section{Results and discussion}
\subsection{Device Design}

\begin{figure}
  \includegraphics[width=10cm]{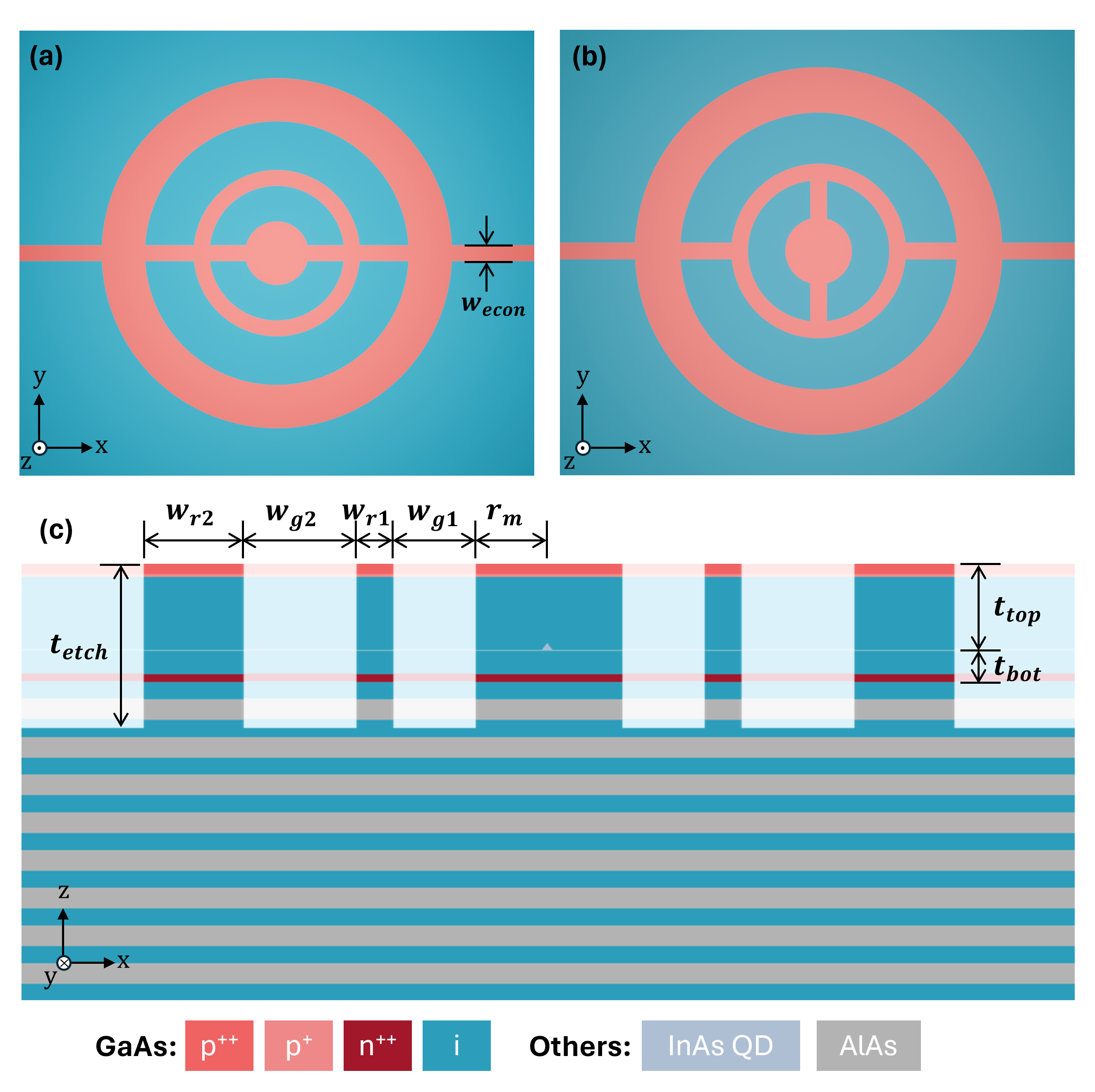}
  \caption{Design schematics of an electrically tunable single-photon source based on a QD integrated into an eCBG in (a) direct-ridge and (b) mazy-ridge configuration. (c) The vertical cross-section of the device illustrates a QD embedded in a PIN diode structure with back-side DBR consisting of 10 $\lambda/4$-thick mirror pairs. The device parameters denoted in the schematics include central mesa radius ($r_m$), ring/gap widths ($w_{r1,r2/g1,g2}$), ridge width ($w_{econ}$), etch depth ($t_{etch}$) and QD top/bottom spacer ($t_{top/bot}$).}
  \label{fgr:device_design}
\end{figure}

Designing a nanophotonic device involves careful consideration of the interplay between the geometry of individual structural elements in the proximity of the photon source (in this case, a QD) and their respective refractive indices. When modifying the well-known optically pumped CBG resonator concept into a ridge-based eCBG, the introduction of contact ridges which can lead to scattering and lateral photon losses is the primary factor influencing, and potentially reducing, the PEE. In this context, two aspects related to the electrical contacts are considered in particular: the electrical contact path length and the width of the ridge.

From the electrical perspective, the ridge should ideally have the shortest path length connecting the central mesa to the external areas to minimize the electrical resistance. This condition constitutes our first proposed design featuring direct-ridge electrical connections, illustrated in Figure~\ref{fgr:device_design}a. In this design, when considering the optical perspective, such a straight ridge could provide an undesired optical loss channel for emitted photons from the central mesa via waveguiding, thereby deteriorating the lateral confinement. In fact, as a countermeasure, labyrinth-like structures have been proposed, where the connections between rings were sectioned and rotated \cite{buchinger_optical_2023}.  As a comparative experimental case study, another design featuring mazy-ridge electrical connections to reduce the effect of lateral waveguides is also proposed here and depicted in Figure~\ref{fgr:device_design}b. To keep the electrical path as short as possible, both designs feature only two rings, and in the mazy-ridge design, only the connections inside the first ring are rotated.

A similar trade-off scenario applies to the width of the ridge as well. For electrical purposes, the ridge width should ideally be as large as possible to minimize the resistance, but as the ridge becomes wider, it could support the propagation of waveguiding modes \cite{schnauber_indistinguishable_2019,hoehne_numerical_2019}, degrading the optical performance of the device. While the optical performance could be predicted and optimized via numerical simulations, the sensible prediction for the electrical behavior of nanoscale structures requires experimental insights, for instance, on the width of depleted layers at the etched surfaces. In fact, our eCBGs are patterned and fabricated utilizing inductively coupled plasma reactive-ion etching (ICP-RIE), which usually results in residual surface defects with deteriorated electrical conductivity \cite{cole_reactive_1992}. As a consequence, a functional ridge must be wide enough to accommodate an electrically active region, sided by defective surfaces. For this, the minimum ridge width was initially selected as 100 nm, and the maximum ridge width was intuitively limited by the effective wavelength of typical InGaAs QD emissions (>900 nm) in GaAs medium, which is about 250 nm. 

Based on the considerations mentioned above, numerical optimizations were performed to identify the optimal device parameters depicted in Figure~\ref{fgr:device_design}c, by applying a Bayesian optimization algorithm to maximize the PEE at the numerical aperture (NA) of 0.81, which is the NA of the optics later used in the experiments. The field distributions and PEE were calculated by a three-dimensional (3D) finite-element method (FEM) in JCMsuite \cite{burger_hp-finite_2015} based on the direct-ridge configuration, exploiting two vertical mirror planes (one along the main ridge axis and one perpendicular to it) to reduce the demanding computational time and resources for full 3D simulations. To sensibly limit the computation time for optimizations, the overall size of the simulated structures was bound by setting search ranges for the mesa radius (\(r_{m}\)), ring widths (\(w_{r1,2}\)), gap widths (\(w_{g1,2}\)), and top/bottom-spacer (\(t_{top/bot}\)) within 100 and 500 nm. The QD was modeled using two point-like dipoles oriented along the x- and y-axis emitting at the wavelength of 930 nm, and the absorption effect of dopants in p- and n-doped GaAs was neglected in the simulation at this wavelength.

\begin{table}[ht]
\caption{Optimized device parameters and simulated PEEs obtained from the optimizations with the minimum \(w_{econ}\) of 100 nm and other parameter search ranges limited to the maximum of 500 nm (except for \(t_{etch}\), which was limited to the maximum of 800 nm). The PEEs (NA = 0.81) were calculated using two perpendicular in-plane (x and y) dipoles oscillating at the wavelength of 930 nm.}
\label{tbl:opt1_devparams_and_PEE}
\begin{tabular}{ccccccc}
\hline
\multirow{2}{*}{\textbf{Structure}} &
  \multicolumn{5}{c}{\textbf{Device parameters {[}nm{]}}} &
  \multirow{2}{*}{\textbf{Sim. PEE {[}\%{]}}} \\ \cline{2-6}
 &
  \textbf{\(w_{econ}\)} &
  \textbf{\(r_{m}\)} &
  \textbf{\(w_{g1,2}\)} &
  \textbf{\(w_{r1,2}\)} &
  \textbf{\(t_{etch}\)} &
   \\ \hline
Direct-ridge& 100 & 243 & 268, 369 & 120, 327 & 630 & 43.2 \\
Mazy-ridge& 100 & 243 & 268, 369 & 120, 327 & 630 & 42.1 \\ \hline
\end{tabular}
\end{table}

The device parameters obtained from the numerical optimization and the calculated PEEs for both designs are listed in Table~\ref{tbl:opt1_devparams_and_PEE}. In this device design, the top spacer (\(t_{top}\)) and bottom spacer (\(t_{bot}\)), which distance the QDs from the surface and the DBR, were determined to be 333 nm and 119 nm, respectively. The uppermost part of the top spacer was chosen as a 50 nm p-doped layer, while the lowermost part of the bottom spacer was chosen as a 30 nm n-doped layer. It is important to point out that the optimal \( w_{econ}\) value for the highest PEE was found at the lower boundary of the defined search domain, which was 100 nm, implying a narrow ridge width was favored. This numerical study, which predicts PEEs of 43.2\% and 42.1\% for the two considered eCBG configurations, confirmed the assumptions mentioned earlier regarding the ridge width and implied that the main limiting criterion for a functional ridge width is the electrical conductivity which needed to be investigated experimentally.

\subsection{Deterministic Fabrication} \label{subsec_deter_fab}

Following the aforementioned optimized design parameters, the heterostructure was epitaxially grown using MBE as follows. Firstly, at a substrate temperature of 605 °C, a 100 nm GaAs buffer layer was deposited on an undoped GaAs (100) wafer, followed by the growth of a DBR consisting of ten pairs of AlAs and GaAs layers with nominal thicknesses of 79 nm and 67 nm, respectively. Subsequently, a 30 nm Si-doped GaAs layer with a doping concentration of $2\times10^{18}$ cm\textsuperscript{-3} was grown at 555 °C. Then, the substrate temperature was returned to 605 °C, and an 86 nm GaAs layer was deposited. Afterward, a 3-minute pause was introduced during which the substrate temperature was lowered from 605 °C to 505 °C, and the arsenic pressure was gradually reduced from $2.25\times10^{-5}$ mbar to $1.50\times10^{-5}$ mbar. After a 2-minute temperature stabilization period, 3 nm of GaAs were deposited. Indium was then deposited in a pulsed mode (4 s deposition, 4 s pause) over 20.5 deposition cycles. During the initial 10.5 cycles, the substrate rotation was halted to create an indium gradient along the [110] direction. The rotation was resumed for the remaining 10 cycles at a speed of 10 rpm. The QDs were then partially capped with 2.6 nm of GaAs at 485 °C. Following this, the temperature was rapidly increased to 60 5°C, and a 280 nm GaAs layer was grown. Lastly, a top gate structure consisting of 10 nm and 40 nm C-doped GaAs layers, with doping concentrations of $1\times10^{18}$ cm\textsuperscript{-3} and $1\times10^{19}$ cm\textsuperscript{-3}, respectively, was deposited at 555 °C. All temperatures mentioned were determined using band-edge thermometry.

Following the epitaxial growth, a sample piece of (5$\times$5) mm\textsuperscript{2} with low QD density (about $2.5\times10^{6}$ cm\textsuperscript{-2}) was selected, on which electrical contacts and \ch{Au} markers were prepared via four EBL steps. The first lithography was performed to pattern areas at the sample corners, which were then etched by 410 nm using ICP-RIE to remove the top p-doped layer and prepare for n-pads. Then, within the etched areas, n-contact pads were patterned and deposited with thicknesses of 10 nm \ch{Ni}, 50 nm Au\textsubscript{0.88}Ge\textsubscript{0.12}, and 40 nm Au. After that, the sample was rapidly annealed at 420 °C for 90 s with a 500 K/min ramping rate under \ch{N2} atmosphere. For the third EBL, p-contacts and markers were patterned, and 10 nm \ch{Ti} and 90 nm \ch{Au} were deposited. Lastly, all p- and n-contacts were deposited with an additional 250 nm \ch{Au}, to achieve the total pad thickness of 350 nm to ensure the durability of the pads during the wire bonding procedure. 

\begin{figure}
  \includegraphics[width=12cm]{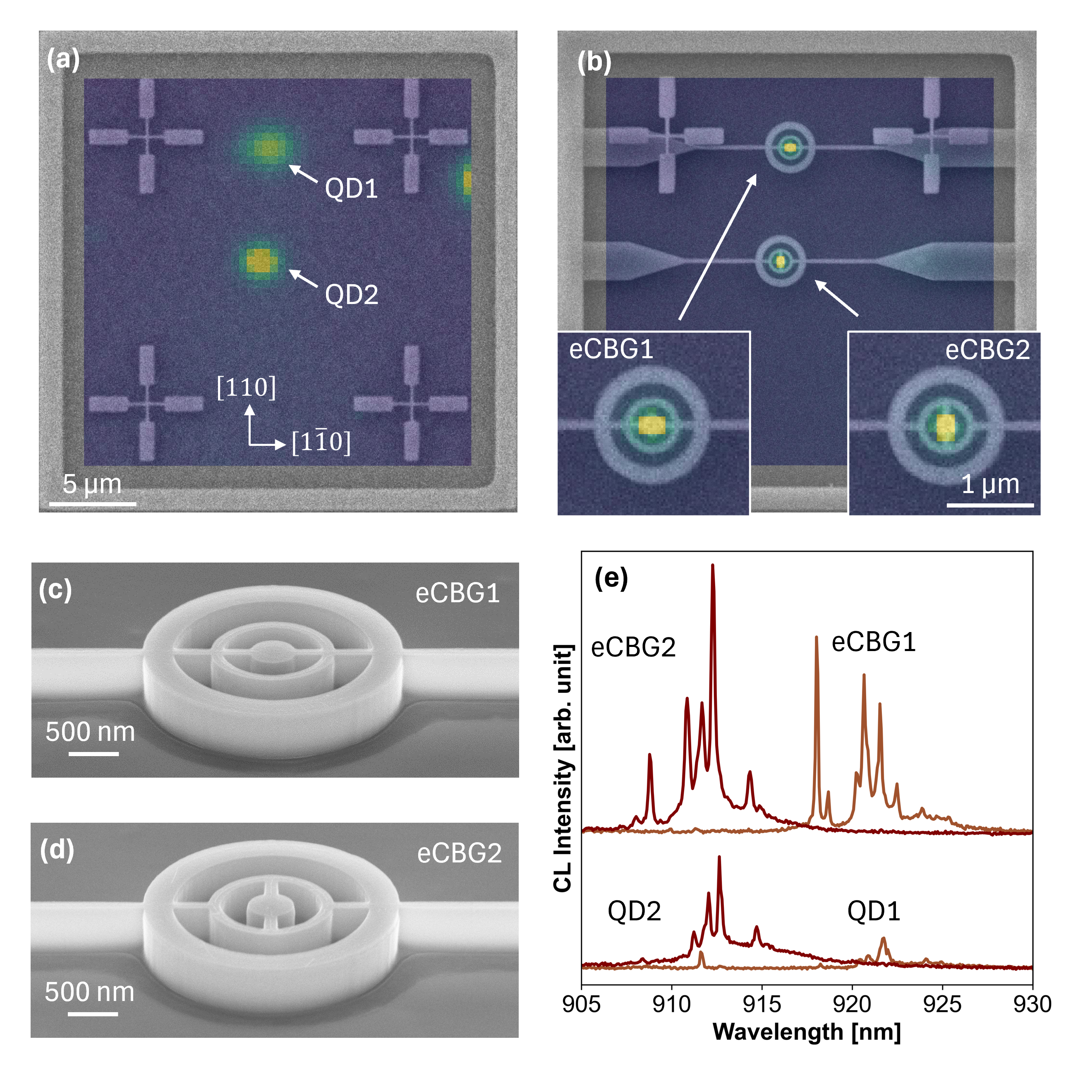}
  \caption{(a) A CL map overlaid on the SEM picture of a marker field, showing two spatially localized  QDs (labeled QD1 and QD2) in a planar structure. (b) A CL map of the same marker field after using marker-based EBL to pattern eCBGs on the QDs deterministically with the main ridge axis aligned along \([1\bar{1}0]\) orientation. QD1 was integrated into a direct-ridge eCBG, whereas QD2 was integrated into a mazy-ridge eCBG. (c-d) SEM images of the fully processed eCBG structures (e) CL spectra of QD1 and QD2 before and after the integration into eCBGs.}
  \label{fgr:device_fabrication}
\end{figure}

Suitable QDs were identified based on their emission wavelength and separation from other QDs on the sample with pre-patterned markers by scanning each marker field using CL mapping at 20 K without applying any bias voltage (open-circuit). A map consisting of an array of CL spectra and a scanning electron microscope (SEM) image, as illustrated using an exemplary field in Figure~\ref{fgr:device_fabrication}a, provides spectral and spatial information about suitable QDs, which can be pre-selected for device fabrication. After identifying the QDs, marker-based EBL \cite{li_scalable_2023} and ICP-RIE were performed to pattern eCBGs at the determined QD positions. As the definitive minimum ridge width for an electrically active eCBG was unknown, the ridge width was intentionally varied from 100 to 130 nm on different patterned structures without changing any other parameters. 

The enhancement from the eCBG integration on the QDs could be observed via a one-to-one comparison by performing another CL mapping on the same marker field after the integration, as depicted in Figure~\ref{fgr:device_fabrication}b, showing that devices were patterned on the determined QDs. Note that the QD emission spectra could be observed on all 14 patterned devices regardless of their patterned ridge widths as they were visible in open-circuit states. In the marker field, two QDs (namely QD1 and QD2) were integrated into a direct-ridge structure (eCBG1) and a mazy-ridge structure (eCBG2) whose SEM images are illustrated in Figure~\ref{fgr:device_fabrication}c and d, respectively. The images indicate smooth etched surfaces and well-defined fabrication conditions. Figure~\ref{fgr:device_fabrication}e shows the CL spectra of both QDs before and after device integration using the same excitation and collection parameters. It is important to note that the intensity between QDs could not be compared fairly as the CL mapping was performed using a parabolic mirror, which resulted in a non-uniform intensity profile throughout the map. Qualitatively, the post-integration spectra of both QDs clearly indicate strongly enhanced extraction efficiency and improved signal-to-noise ratio. 

\subsection{Electric-field Dependent Photoluminescence Measurements}
\begin{figure}
  \includegraphics[width=11cm]{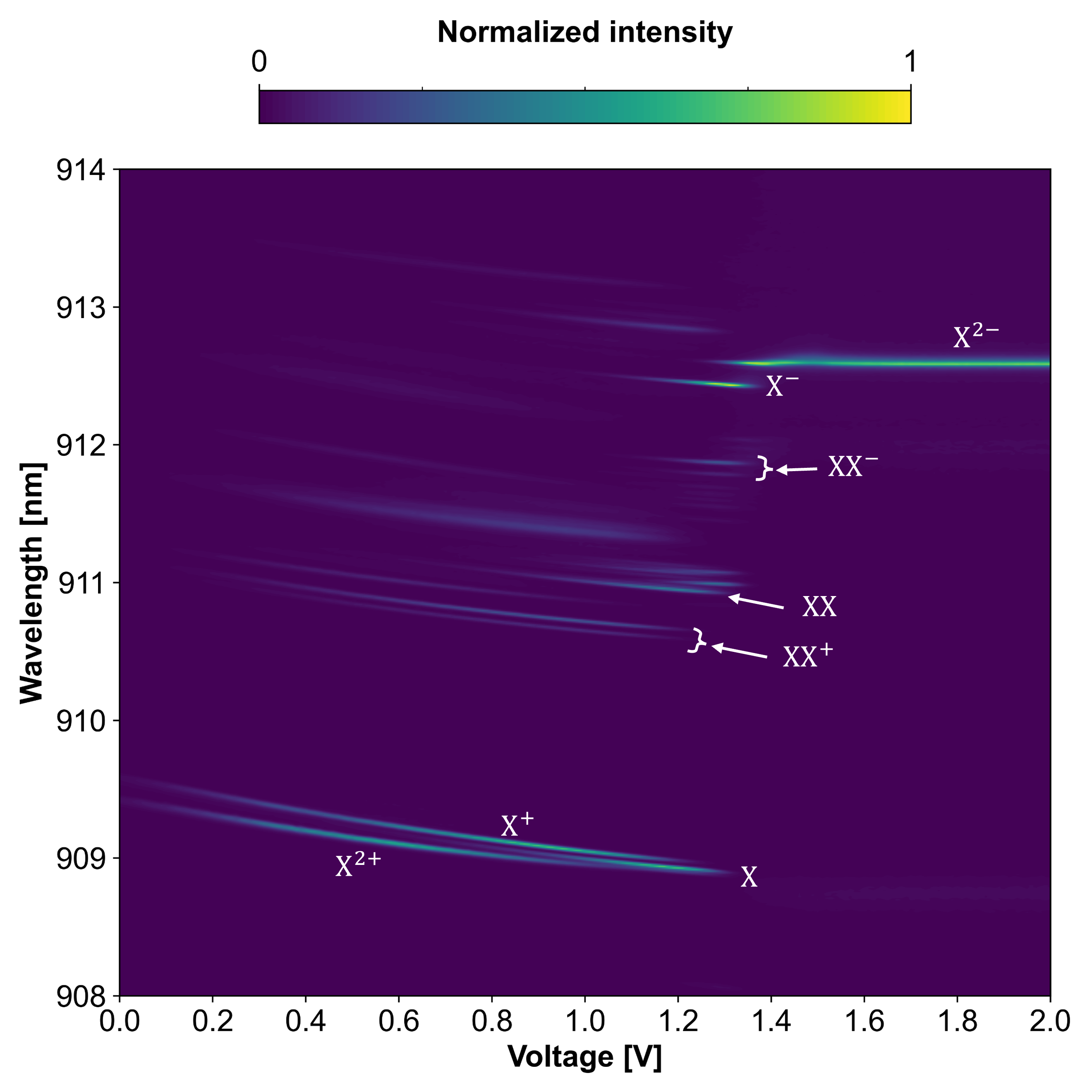}
  \caption{Contour plot of bias-voltage dependent $\mu$PL spectra of the eCBG2-QD recorded at 4 K showing several emission lines originating from excitonic ($\mathrm{X}$), biexcitonic ($\mathrm{XX}$), singly-charged ($\mathrm{X}^{+/-}$, $\mathrm{XX}^{+/-}$), and doubly charged ($\mathrm{X}^{2+/2-}$) states observed at different bias voltages. The lines observed at bias voltages under 1.4 V exhibit spectral shifts due to the quantum-confined Stark effect, allowing for emission wavelength tuning via external electrical controls.}
  \label{fgr:vols_PL}
\end{figure}

After the deterministic fabrication, electrical connections between an external voltage source and the diode structures of the fabricated devices were made through wire bonding between a chip carrier and deposited metallic contact pads. The chip carrier was then installed in the optical setup described in Section~\ref{subsec_optical_setup} for bias-voltage dependent $\mu$PL measurements. The eCBGs were investigated under pulsed (80 MHz) ps-mode excitation via wetting layer at 865 nm using a Ti:Sapphire laser. By varying the external bias voltage (\(V_{ex}\)) between 0 and 3 V, we observed that the fabricated direct-ridge eCBGs whose ridges were narrower than (110$\pm$5) nm exhibited no voltage dependency on their optical spectra, as well as the mazy-ridge eCBGs whose ridges were narrower than (120$\pm$5) nm. Both eCBG1 and eCBG2, as well as other eCBGs that consisted of wider ridges than these values, showed clear and similar bias-voltage dependency, as illustrated using eCBG2 as an example in Figure~\ref{fgr:vols_PL}.

The $\mu$PL contour plot displays different QD emission lines observed at different \(V_{ex}\). Most lines, which appeared below about 1.4 V, featured the quantum-confined Stark effect by which the emission wavelengths shifted gradually over the \(V_{ex}\) change, whereas the emission lines appearing above 1.4 V remained energetically constant. This behavior can be explained by the electronic band-bending of the PIN diode structure, which was initially caused by the built-in electric field (\(E_{in}\)) \cite{miller_band-edge_1984}. At zero bias, the electronic bands were strongly bent by the field which promotes the quantum tunneling rate of electrons from the wetting layer to GaAs, reducing carrier capturing into the QD, hence, the low PL emission rates were observed at 0 V \cite{fry_electric-field-dependent_2000,oulton_manipulation_2002}. As the \(V_{ex}\) increased, the band-bending and electron tunneling weakened, enabling the carrier capturing in the QD and subsequent carrier recombination, hence the emission lines became visible while shifting energetically \cite{oulton_manipulation_2002,bennett_giant_2010}. Notably, bias-induced energetic shifts of up to 0.7 meV were observed in the studied devices. When the externally applied electric field became as equally strong as the \(E_{in}\), which was at around \(V_{ex}\) = 1.4 V), flat-band conditions were achieved, and no quantum confined Stark tuning could be observed anymore.  

\begin{figure}
\includegraphics[width=16cm]{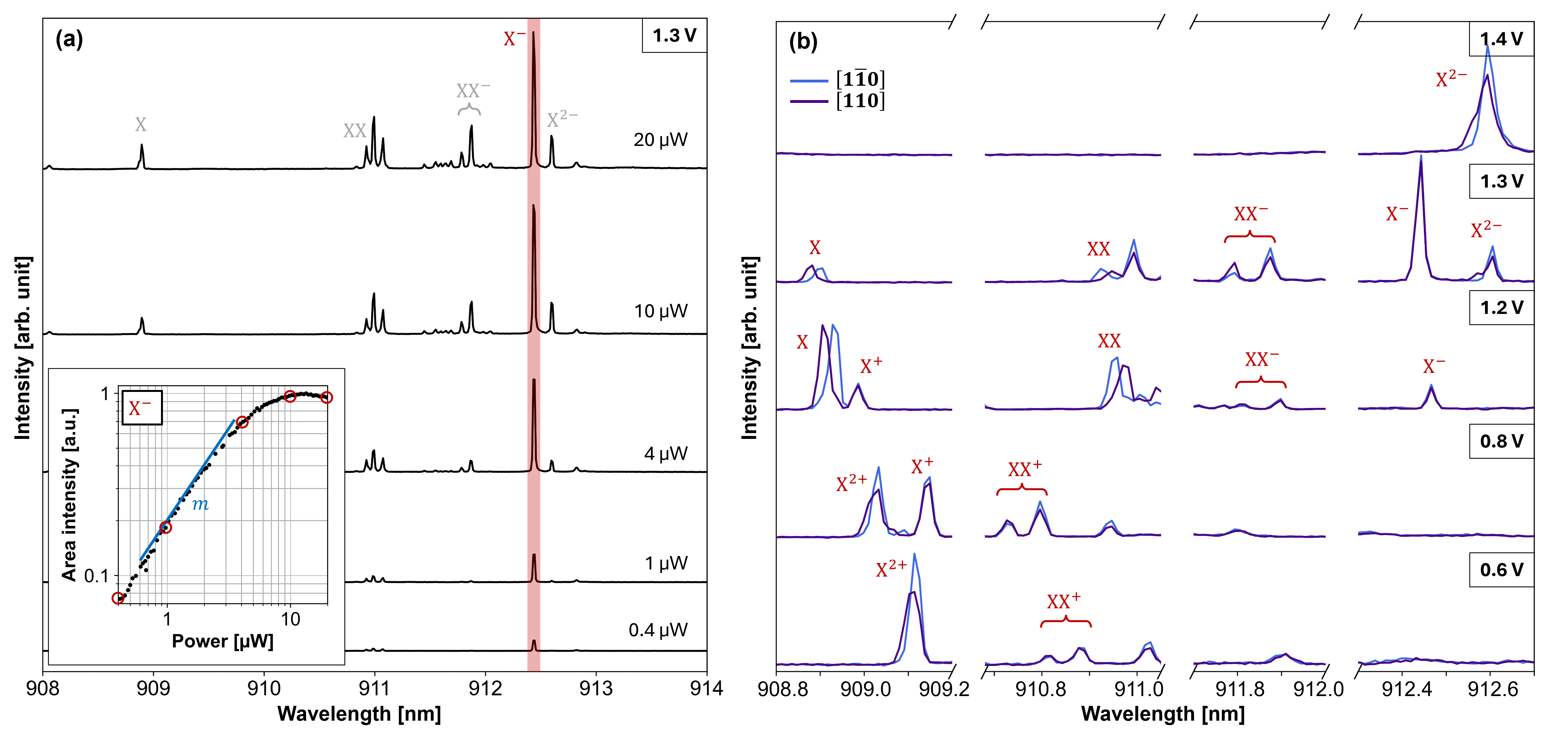}
\caption{(a) Power-dependent $\mu$PL spectra from eCGB2 biased at 1.3 V, and the area intensity of the $\mathrm{X}^{-}$ emission line plotted against excitation power in a log-log scale. (b) Polarization-dependent $\mu$PL spectra from eCBG2 biased at different bias voltages, showing multiple emission lines. The two perpendicular polarization angles plotted correspond to the \([1\bar{1}0]\) (blue) and [110] (purple) orientations of the sample.}
  \label{fgr:pows_pols_PL}
\end{figure}

The nature of observed emission lines was determined by performing power- and polarization-dependent $\mu$PL measurements at the corresponding \(V_{ex}\). For the power-dependent measurements, the $\mu$PL spectra were recorded at different optical excitation power. By selectively fitting the area intensity of each emission line and plotting them against the excitation power in a log-log scale, a power exponent ($m$) can be extracted from the slope of the linear section in the plot. Figure~\ref{fgr:pows_pols_PL}a shows an example of a power-dependent $\mu$PL measurement at 1.3 V where an emission line (later determined as the  $\mathrm{X}^{-}$ emission line) was selectively fitted. The same process was done for other lines and the findings are listed in Table~\ref{tbl:vols_pows_pols}. With extracted $m$'s, excitonic and biexcitonic lines were categorized by their linear ($m\sim1$) and superlinear ($m$>1) power dependencies, respectively \cite{finley_charged_2001}. Complementarily, the information on polarization properties of these emission lines was necessary to determine their origins unambiguously, caused by the spin-related fine structures \cite{akimov_electron-hole_2005,ediger_fine_2007,warming_hole-hole_2009}. For this, the polarization-dependent measurements were performed by inserting a rotatable $\lambda/2$-plate and a linear polarizer in the collection path of the $\mu$PL setup. By changing the $\lambda/2$-plate angle, different polarization components of each emission line were recorded. Figure~\ref{fgr:pows_pols_PL}b shows the two polarization components of each emission line along \([1\bar{1}0]\) and [110] orientations of the sample. For each line, the peak energy difference between the two polarization angles and the linewidth (full-width half-maximum measured at a certain polarization angle where the narrowest emission was observed) were determined and are also listed in Table~\ref{tbl:vols_pows_pols}. Combining the findings from these two measurements, the emission lines were determined as follows.

\begin{table}[ht]
\caption{Emission lines observed from eCBG2 at different bias voltages and their relevant parameters, including emission wavelengths ($\lambda$) and central energies (E), power exponents ($m$) from power-dependent $\mu$PL measurements, fitted peak energy deviations ($\Delta$E) and minimum emission linewidths (FWHM) from polarization-dependent $\mu$PL measurements.}
\label{tbl:vols_pows_pols}
\begin{tabular}{ccccccc}
\hline
Emission & Voltage {[}V{]} & $\lambda$ {[}nm{]} & E {[}meV{]} & $m$ & $\Delta$E {[}$\mu$eV{]} & FWHM {[}$\mu$eV{]} \\
\hline
$\mathrm{X}^{2+}$     & 0.6 & 909.11 & 1366.58 & 1.51±0.01 & 9.0±1.8& 31.5±2.9\\
$\mathrm{X}^{+}$      & 0.8 & 909.14 & 1366.53 & 0.85±0.01 & None     & 23.7±2.5\\
$\mathrm{XX}^{+}_{1}$ & 0.8 & 910.73 & 1364.15 & 1.50±0.02 & None     & 24.8±4.6\\
$\mathrm{XX}^{+}_{2}$ & 0.8 & 910.80  & 1364.05 & 1.60±0.03 & None     & 29.7±3.1\\
X                     & 1.2 & 908.92 & 1366.87 & 0.82±0.01 & 37.2±2.6& 24.6±4.7\\
XX                    & 1.2 & 910.96 & 1363.81 & 1.28±0.01 & 34.4±2.4& 27.2±2.4\\
$\mathrm{XX}^{-}_{1}$ & 1.3 & 911.79& 1362.57& 1.69±0.04 & None     & 28.9±3.1\\
$\mathrm{XX}^{-}_{2}$ & 1.3 & 911.87& 1362.44& 1.78±0.02 & None     & 29.4±1.2\\
$\mathrm{X}^{-}$      & 1.3 & 912.44 & 1361.60 & 0.97±0.01 & None     & 16.6±1.1\\
$\mathrm{X}^{2-}$     & 1.4 & 912.59 & 1361.37 & 1.29±0.01 & 7.7±1.4& 46.0±1.2\\
\hline
\end{tabular}
\end{table}

Firstly, the most fundamental states of QD, exciton-biexciton (X-XX) emissions were observed at around 1.2 V exhibiting a strong polarization dependency with the opposite energetic deviation direction. The energy deviations indicate the X-XX fine-structure splittings of (37.2$\pm$2.6) for X and (34.4$\pm$2.4) $\mu$eV for XX, respectively, which are typical values for an InGaAs QD \cite{seguin_size-dependent_2005}. Secondly, when varying \(V_{ex}\) slightly to a lower value (0.8 V), three prominent lines were observed without any polarization-dependent energetic variation, consisting of a single line with  $m\sim1$ and double lines with $m>1$. This suggests that they originate from singly-charged excitonic and biexcitonic states. Moreover, the double lines exhibited clear polarization-dependent intensities, where each line was more prominent than the other at different polarization angles. This strongly supports that they originate from the relaxation of the charged biexcitonic state into the two triplet states of excited-trion, which typically generate elliptically-polarized photons \cite{akimov_electron-hole_2005,kavokin_fine_2003}. Thirdly, when decreasing \(V_{ex}\) even further (<0.8 V), a new prominent line appeared having $m>1$ and showing a polarization-dependent emission. The emission was observed as a single line at a certain polarization angle, but the line broadened with a lower peak intensity when observed at different polarization angles. This suggests that the emission consists of two different lines, one without polarization dependency and another with polarization-dependent energetic variation, which could be attributed to the two triplet transitions from doubly-charged excitonic states \cite{urbaszek_fine_2003,ediger_fine_2007}. 

Another similar set of lines was observed as well when varying the voltage to the higher values (\(V_{ex}\)>1.2 V), indicating that they originate from the analogous spin states with the opposite charge sign. Note that, at lower voltages (\(V_{ex}\)<1.2 V) where the band-bending exists, the tunneling rate of electrons is more significantly promoted than that of holes, due to the smaller effective mass, suggesting the hole is more effectively captured into the QD and the prominent emission lines are expected to originate from the positively charged states \cite{fry_electric-field-dependent_2000,oulton_manipulation_2002}. On the other hand, at higher voltages (\(V_{ex}\)>1.2 V) where the tunneling weakens, the electron capturing in the QD becomes effective, suggesting that the emission lines in this regime originate from negatively charged states instead. Additionally, the redshifts of the higher negatively-charged emission lines also support the assumptions, in agreement with previously reported studies \cite{urbaszek_fine_2003,baier_optical_2001,warburton_optical_2000}. Consequently, all the prominent emission lines in the voltage-dependent $\mu$PL spectra were assigned, ranging from the positive doubly-charged line ($\mathrm{X}^{2+}$) at low voltages to the negative doubly-charged line ($\mathrm{X}^{2-}$) at high voltages, as shown in Figure~\ref{fgr:vols_PL}.

Understanding the spin-related origins of the emission lines is crucial for application in, for instance, quantum repeater networks, which require well-defined light-matter interactions between the flying and stationary qubits \cite{heindel_quantum_2023}. The results presented in this section demonstrate the determination of the origins of multiple QD emission lines and confirm the capability of the designed eCBG devices, allowing for deterministic controls of electronic states of QDs in the cavity. This feature can significantly enhance the performance of spin-photon interfaces, crucial for mediating photon entanglements in quantum repeater networks \cite{luo_spinphoton_2019}. The emission wavelength tunability of the QD in the cavity enabled by this design can also benefit the applications with quantum memories \cite{neuwirth_quantum_2021}. Moreover, the design can be directly applied to QD molecules, which have promising applications in photonic cluster-state generation \cite{azuma_all-photonic_2015,schall_bright_2021,bopp_quantum_2022,vezvaee_deterministic_2022}. 

\subsection{Optical Enhancements}
\begin{table}[ht]
\caption{Structural parameters of fabricated device obtained from SEM imaging (with measurement errors $\leq$ 5 nm) and the simulated and measured PEEs (NA = 0.81). The simulations were performed using two perpendicular in-plane (x and y) dipoles oscillating at the corresponding wavelengths to the measured emissions. For planar QDs, the PEE was simulated for the wavelength range of 900 -- 930 nm, and measured from nine different QDs.}
\label{tbl:exp_devparams_PEE}
\begin{tabular}{ccccccccc}
\hline
\multirow{2}{*}{\textbf{Structure}} &
  \multicolumn{5}{c}{\textbf{Device parameters {[}nm{]}}} &
  \multicolumn{2}{c}{\textbf{PEE {[}\%{]}}} & \multirow{2}{*}{\textbf{$\lambda$ {[}nm{]}}}\\ \cline{2-8} 
 &
  \textbf{\(w_{econ}\)} &
  \textbf{\(r_{m}\)} &
  \textbf{\(w_{g1,2}\)} &
  \textbf{\(w_{r1,2}\)} &
  \textbf{\(t_{etch}\)} &
  \textbf{Simulated} &
  \textbf{Measured}  &\\ \hline
Planar      & -   & -   & -        & -        & 0   & 2.02$\pm$0.19 & 1.69$\pm$0.59  &900 -- 930\\
eCBG1& 122 & 240 & 250, 368 & 137, 324 & 633 & 24.53& 12.6$\pm$1.4   &918.1\\
eCBG2& 132 & 245 & 240, 368 & 137, 328 & 633 & 31.93& 30.4$\pm$3.4   &912.7\\ \hline
\end{tabular}
\end{table}

As mentioned in the previous section, the electrically functioning eCBGs were experimentally found to require wider ridges than the originally optimized value (100 nm). For a fair comparison between the simulation and experimental results, the actual geometry of the fabricated structures (eCBG1 and eCBG2) was measured using SEM, and the PEEs were re-simulated based on the measured structural parameters which are listed in Table~\ref{tbl:exp_devparams_PEE}. Note that the experimental PEE can deviate from the simulated value due to the possible mismatch of the QD position and the integrated eCBG \cite{madigawa_assessing_2024}, which could not be determined in this work, limited by the small diameter of the central mesa.    

To obtain experimental PEEs, the $\mu$PL setup efficiency was measured using a continuous-wave laser at the same wavelength as the QD emission, yielding the value of (7.3$\pm$0.8)\%. Then, nine different QDs in the planar structure and the studied devices (eCBG1 and eCBG2) were electrically biased at 1.5 V to achieve the flat-band conditions for maximum electric-field dependent PL emissions \cite{fry_electric-field-dependent_2000}, and the emission spectrum from each QD was recorded under pulsed (80 MHz) excitation at 800 nm using a Ti:Sapphire laser with the corresponding saturation optical pump power. The PEE was calculated using the formula: \[\mathrm{PEE}=\frac{n}{\eta_{setup}*f}\]where $n$ is the single-photon emission rate, $\eta_{setup}$ is the setup efficiency, and $f$ is the excitation frequency. For eCBG1 and eCBG2, single-photon emission rates were measured to be (732$\pm$18) kHz and (1.77$\pm$0.34) MHz, equivalent to PEEs of (12.6$\pm$1.4)\%. and (30.4$\pm$3.4)\%, respectively. For planar QDs, the average (and standard deviation) emission rate was found to be (98.6$\pm$26.1) kHz, which led to an average PEE of (1.69$\pm$0.59)\%.

The comparison between the simulated and experimental PEE for eCBG2 shows impressive conformity, indicating accurate numerical modeling and good fabrication quality. The reliability of this comparison can be supported by the results from planar QDs, which also show good agreement between the simulation and the measured value. Deviated from the simulation, eCBG1 yielded noticeably lower efficiency, similar to the other two fabricated direct-ridge eCBGs with the same nominal parameters. The cause for the larger deviation of the experimental PEE from the simulated one for eCBG1 than that of eCBG2 may be multi-factored. Firstly, although all eCBGs were fabricated under identical conditions, since the actual eCBG-to-QD position offsets could not be determined, the effect of alignment mismatch cannot be completely ruled out. Secondly, the simulations were performed with a QD modeled as two equally contributing perpendicular point-like dipoles, which might not represent the actual QD emission well enough in the presence of ridge-like structures breaking the cylindrical symmetry. As a result, the simulation may have underestimated the undesired direct-ridge effect, causing a larger difference between simulated and experimental PEEs only for direct-ridge eCBGs. An additional finding described as follows supports this assumption further. 

\begin{figure}
  \includegraphics[width=13cm]{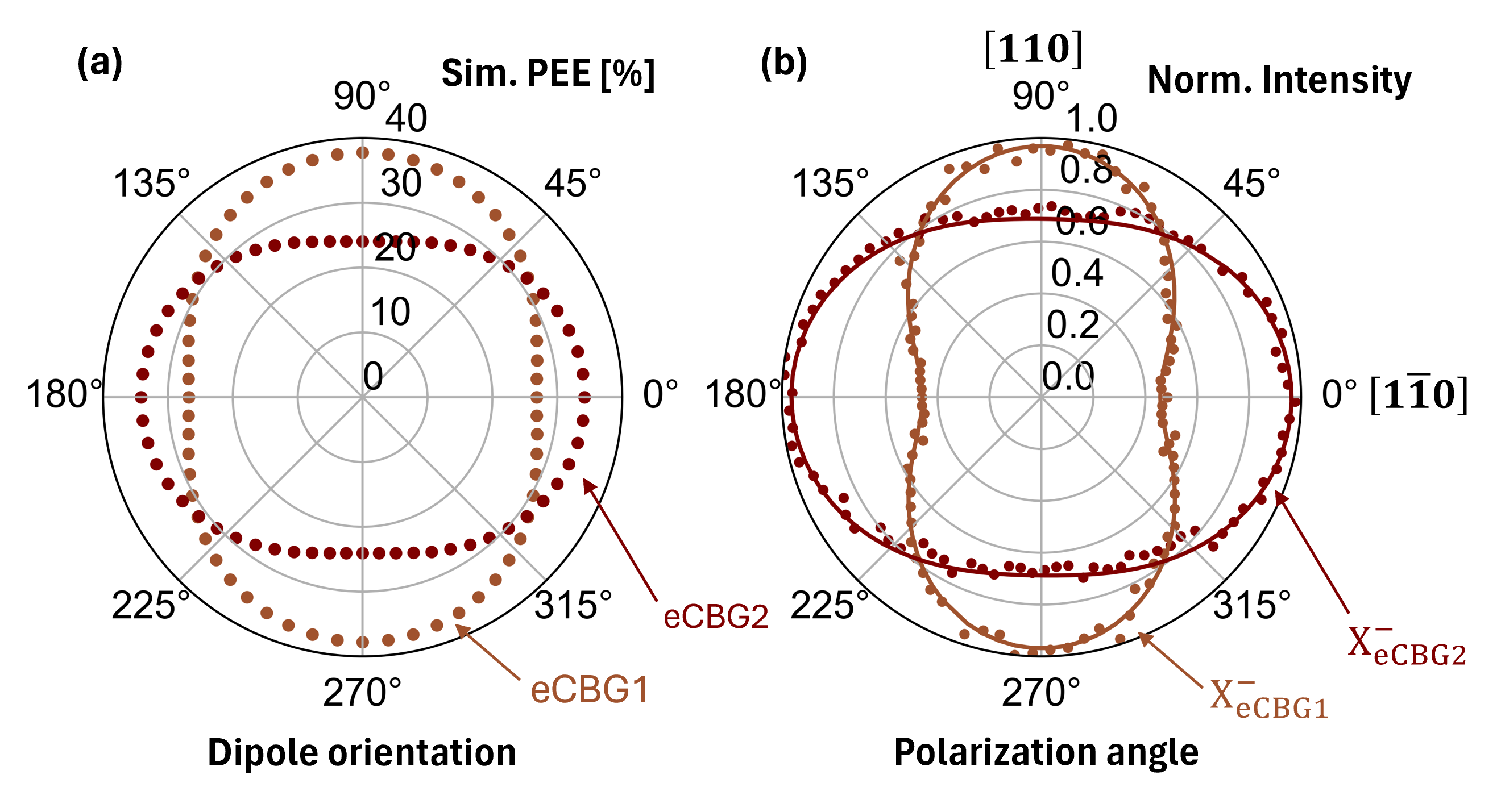}
  \caption{(a) Simulated PEEs (NA = 0.81) of eCBG1 and eCBG2 using a single point-like in-plane dipole source oriented at different angles related to the main ridge axis. (b) Normalized intensities of the $\mathrm{X}^{-}$ emission lines from eCBG1 and eCBG2 in polarization-dependent PL measurements, where the 0° and 90° correspond to the linear polarization components along \([110]\) and \([1\bar{1}0]\) orientations, respectively.}
  \label{fgr:fitpols_PL}
\end{figure}

Due to the presence of the ridges in eCBGs, the polarization-dependent optical performance of the fabricated devices was intuitively anticipated. On the simulation side, the effect was predicted by calculating the PEE using a single in-plane dipole oriented at different angles related to the main ridge axis. The simulation results for eCBG1 and eCBG2, plotted in Figure~\ref{fgr:fitpols_PL}a, illustrate good enhancements for the dipoles oscillating perpendicularly to the innermost ridge axis, whereas the enhancements deteriorated the most when the dipole was placed in parallel instead. The simulated PEEs could be normalized and calculated into a degree of linear polarization: $DLP = (I_{max}-I_{min})/(I_{max}+I_{min})$, where \(I_{max/min}\) are maximum and minimum intensity (equivalent to PEE). Interestingly, both eCBG1 and eCBG2 yielded the same simulated DLP of 17\%, regardless of the difference in the ridge continuity. On the experimental side, the effect could be observed in polarization-dependent $\mu$PL measurements mentioned in the previous section. To exemplify the effect, the $\mathrm{X}^{-}$ emission line (912.44 nm) which is intrinsically circularly polarized was selected, and the fitted polarization-dependent intensity plots are shown in Figure~\ref{fgr:fitpols_PL}b. Note that the polarization angle of 0° corresponded to the polarization components along the \([1\bar{1}0]\) orientation of the sample, to which the main ridge axes of the fabricated devices were aligned. Qualitatively, the experimental plots closely match the simulation results well, clearly confirming the polarization-dependency of PEEs. The experimental DLPs were calculated in the same fashion as earlier, yielding (36$\pm$1)\% and (17$\pm$1)\% for eCBG1 and eCBG2, respectively. For eCBG1, the experimental result implies poorer enhancement of the emission along the main ridge axis than the simulation predicted, which was another piece of evidence signifying the underestimation of the direct-ridge effect on the optical enhancement of the eCBG. On the other hand, another perfect match between simulation and experimental results on eCBG2 could be observed for DLP values. Thus, we conclude that this polarization-dependency of PEEs is mostly affected by the innermost ridges, which suggests that the outer ridges can be modified (e.g. for larger widths) to support electrical conductivity without significantly degrading the optical property.

\begin{table}[ht]
\caption{Optimized device parameters and simulated PEEs obtained from the optimizations with the minimum \(w_{econ}\) of 120 nm and other parameter search ranges up to the maximum of 1000 nm. The PEEs (NA = 0.81) were calculated using two perpendicular in-plane (x and y) dipoles oscillating at the wavelength of 930 nm.}
\label{tbl:opt2_devparams_and_PEE}
\begin{tabular}{ccccccc}
\hline
\multirow{2}{*}{\textbf{Structure}} &
  \multicolumn{5}{c}{\textbf{Device parameters {[}nm{]}}} &
  \multirow{2}{*}{\textbf{Sim. PEE {[}\%{]}}} \\ \cline{2-6}
 &
  \textbf{\(w_{econ}\)} &
  \textbf{\(r_{m}\)} &
  \textbf{\(w_{g1,2}\)} &
  \textbf{\(w_{r1,2}\)} &
  \textbf{\(t_{etch}\)} &
   \\ \hline
Direct-ridge& 120 & 256 & 960, 607 & 257, 628 & 630 & 55.65\\
Mazy-ridge& 120 & 256 & 960, 607 & 257, 628 & 630 & 56.54\\ \hline
\end{tabular}
\end{table}

To demonstrate the capability of improving the current designs, another numerical optimization was performed by setting \(w_{econ}\) to 120 nm and expanding the parameter search ranges for ring and gap widths up to 1000 nm. The newly optimized parameters yielded the PEE > 50\%, listed in Table~\ref{tbl:opt2_devparams_and_PEE}, promising a substantial improvement without changing the heterostructure parameters. With numerical refinement, the proposed design can be adapted with more pairs of back-side DBR to boost the PEE even further. 

\subsection{Single-photon Characteristics}
\begin{figure}
  \includegraphics[width=16cm]{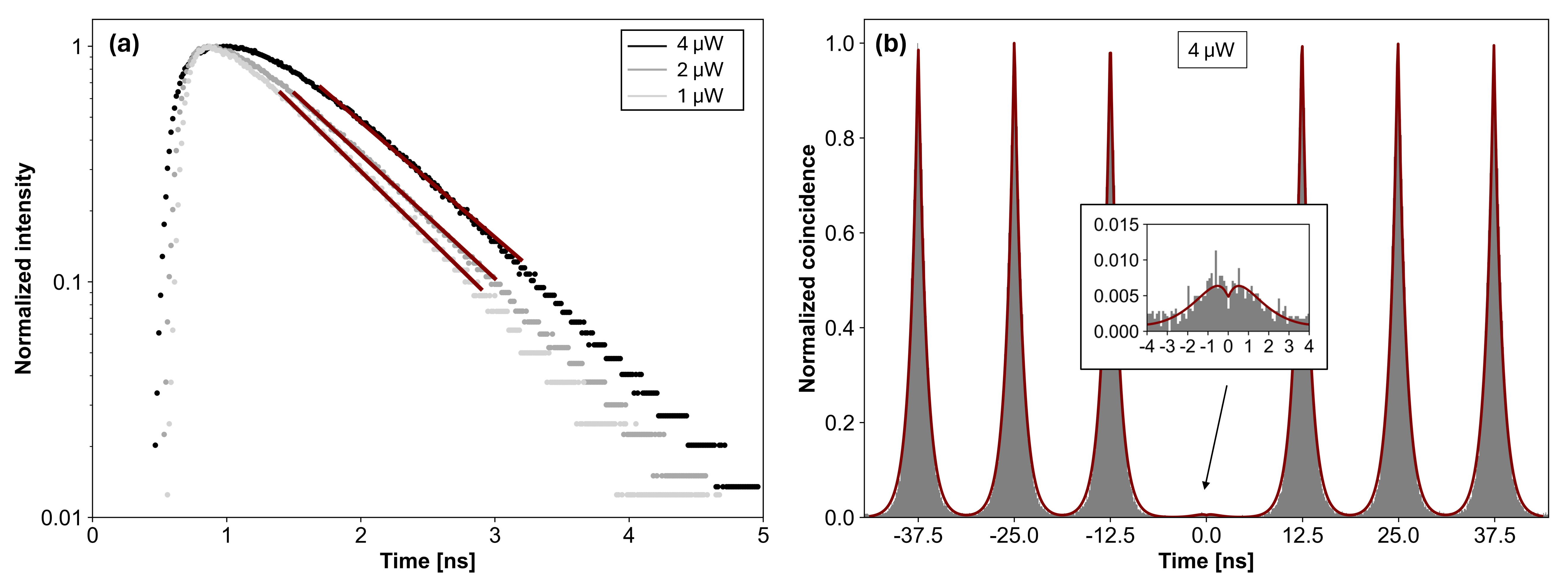}
  \caption{(a) Time-resolved $\mu$PL measurements of the $\mathrm{X}^{-}$ emission line from eCBG2 biased at 1.3 V and measured at 4 K under 1, 2, and 4 $\mu$W excitation powers, which correspond to the emissions with $\sim$ 20$\%$, 40$\%$, and 70$\%$ of the saturation intensity, respectively. The respective fitted profiles (red lines) yield decay lifetimes of 0.78, 0.83, and 0.89 ns (with fitting errors < 0.01 ns). (b) The Hanbury Brown and Twiss correlation statistics $g^{(2)}(\tau)$ of the same emission line under 4 $\mu$W excitation power. The histogram shows excellent multi-photon suppression $g^{(2)}(0)$ of (1.2±0.2)\%, calculated from the area under the fitted profile (red line), and a post-selected $g^{(2)}(0)$ of (0.5±0.3)\%.} 
  \label{fgr:LT_HBT}
\end{figure}

In addition to the optical performance of the fabricated devices, the quantum optical properties were also assessed by performing time-resolved $\mu$PL measurements. For this, the $\mathrm{X}^{-}$ emission line (912.44 nm) from eCBG2 was selected, as it appeared near the flat-band condition, resulting in a bright emission and a resolution-limited linewidth of (16.6$\pm$1.1) $\mu$eV. The experiment was conducted under pulsed (80 MHz) ps-mode excitation at 865 nm using a Ti:Sapphire laser, and the detection path included a monochromator as a spectral filter and a photon correlation setup equipped with superconducting nanowire single-photon detectors (SNSPDs). Firstly, the time-resolved $\mu$PL measurements were performed under different excitation powers, and the results are illustrated in Figure~\ref{fgr:LT_HBT}. Note that excitation powers of 1, 2, and 4 $\mu$W corresponded to the emissions with $\sim$ 20$\%$, 40$\%$, and 70$\%$ of the saturation intensity, respectively, as previously shown in Figure~\ref{fgr:pows_pols_PL}a. The emission lifetimes were found to increase gradually from (0.78$\pm$0.01) to (0.89$\pm$0.01) ns, as the power increased from 1 to 4 $\mu$W. For further measurements, the excitation power of 4 $\mu$W was applied to maximize the emission rate. 

Following the lifetime measurements, a Hanbury Brown and Twiss (HBT) setup \cite{hanbury_brown_question_1956} was used to assess the single-photon purity. The recorded photon correlation histogram, plotted in Figure~\ref{fgr:LT_HBT}, shows a very low coincidence rate observed near the zero-time delay, which can be expected from carrier-recapturing processes due to the non-resonant excitation scheme. Nevertheless, the emission features an excellent multi-photon suppression rate with a $g^{(2)}(0)$ of (1.2$\pm$0.2)\% and a post-selected $g^{(2)}(0)$ of (0.5$\pm$0.3)\%, which are equivalent to the non-post-selected and post-selected single-photon purities of (98.8$\pm$0.2)\% and (99.5$\pm$0.3)\%, respectively. Notably, the multi-photon suppression rate can be further improved by either reducing the excitation power, at the cost of a lower emission rate or by operating the device under resonant excitation \cite{ding_-demand_2016}. 

\begin{figure}
  \includegraphics[width=10cm]{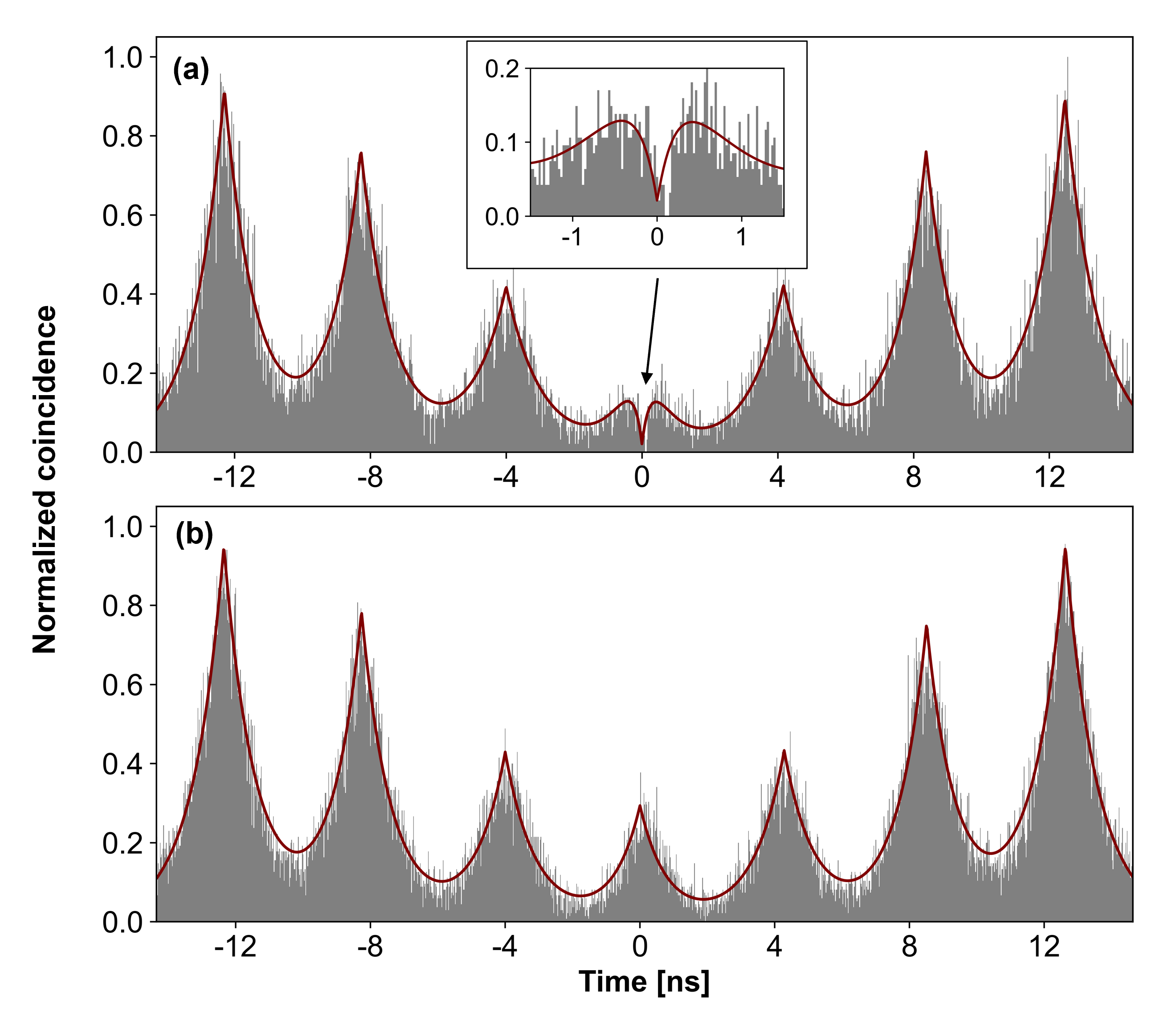}
  \caption{The Hong-Ou-Mandel correlation statistics $g^{(2)}(\tau)$ of photons emitted from the $\mathrm{X}^{-}$ emission line of eCBG2 biased at 1.3 V and measured at 4 K, for parallel (a) and orthogonal (b) polarization. The non-post-selected visibility ($V_{raw}$) of (25.8±2.1)\% and the post-selected visibility ($V_{post}$) of (92.8±4.8)\% were calculated.}
  \label{fgr:HOM}
\end{figure}

After determining the single-photon purity, the photon indistinguishability was assessed using a Hong-Ou-Mandel (HOM) correlation setup \cite{hong_measurement_1987}. In this case, a sequence of double excitation pulses with a 4 ns pulse separation was used, which was matched with a 4 ns fiber delay line in one interferometer arm in the HOM setup. The HOM interference histograms from parallel and orthogonal polarization configurations are plotted in Figure~\ref{fgr:HOM}. The non-ideal raw visibility was obtained as $V_{raw}$ = (25.8$\pm$2.1)\%, evidenced by observed coincidences near the zero-time delay. These coincidences were observed, potentially because the QD had a non-enhanced recombination lifetime allowing the carriers to interact with the solid-state environment, and due to spectral diffusion induced by charge fluctuations at the etched surface near the QD \cite{manna_surface_2020}. However, prominent indistinguishability could be observed at zero-time delay in the parallel polarization configuration, leading to the post-selected visibility $V_{post}$ = (92.8$\pm$4.8)\%. This indicates that QD could intrinsically emit indistinguishable photons, suggesting that optimizing the design toward Purcell enhancement, as well as operating the device under resonant excitation and applying surface passivation, could improve the raw visibility of the device further. 

\section{Conclusion}

In this report, we proposed and implemented an innovative concept for high-performance electrically controlled QD-CBG single-photon sources. By going beyond simple optically controlled structures, our work can   To maximize the brightness of the quantum devices we explored two different ridge-based eCBG designs, based on the trade-off between electrical and optical performance. The sensitivity of the electrical path length and the ridge width to the performance prompted careful optimizations between numerical and experimental results. A direct-ridge configuration featured better electrical connection, requiring a slightly narrower minimum ridge width for an electrically functioning device, while optically performing poorer, especially for the polarization in parallel to the ridge axis. On the other hand, a mazy-ridge configuration featured a slightly longer electrical path length, requiring a wider minimum ridge width to function, but yielded superior optical enhancement. The determinsitically fabricated mazy-ridge eCBG-QD achieved a PEE of (30.4±3.4)\%, in quantitative agreement with numerical simulations which promise substantial improvement with PEE >50 \% when using re-optimized parameters based on the present experiment findings. The electrical operation of the eCBG-QD yields precise charge control and a quantum-confined Stark effect, providing robust capability to fine-tune the emission wavelength up to 0.7 nm and selectively enhance emission lines from specific QD states. The single-photon emission characteristics show impressive a single-photon purity of (98.8±0.2)\% and post-selected visibility of (92.8±4.8)\%. The results underscore the high potential of our advanced quantum device concept, which can pave the way for many applications of photonic quantum information technology requiring not only excellent optical and quantum optical properties but also tight electrical control of the integrated quantum emitters.    

\section{Experimental methods}

\subsection{Optical Setups} \label{subsec_optical_setup}
The $\mu$PL measurements were performed on the sample cooled to 4 K in a closed-cycle cryostat using a Ti:Sapphire laser operating in pulsed ps-mode (80 MHz) at tunable wavelengths (800-865 nm). The emitted photons were collected with a cold aspheric objective lens (NA = 0.81). With a beam splitter, which separates excitation and collection paths, the collected beam was directed through a 900 nm long-pass filter to a monochromator with a 1500-line-per-nm grating and detected using a CCD camera. For time-resolved measurements, two different correlation setups, equipped with SNSPDs, were used: (1) HBT setup, where the spectrally filtered photons were sent through a 1:1 single-mode fiber beam splitter and detected with two SNSPDs were recorded, and (2) Hong-Ou-Mandel setup, where the spectrally filtered photons were sent through a polarization-maintaining beam splitter in combination with polarization-control paths (with $\lambda/2-$ and $\lambda/4-$plates) and a $\sim$4 ns time delay on one arm, before sent back into a 1:1 beam splitter for two-photon interference effect and detected with two SNSPDs. In these correlation setups, the time correlation between two detected photons in each detector was recorded.

\begin{acknowledgement}
This work was financially supported by the German Federal Ministry of Education and Research (BMBF) through the project “QR.X – Quantenrepeater.Link” with funding IDs: 16KISQ012 and 16KISQ014. The authors further thank Johannes Schall, Ching-Wen Shih, Chirag Palekar, Lucas Rickert, Niels Heermeier, and Aris Koulas-Simos for technical support and guidance. 
\end{acknowledgement}

\begin{suppinfo}
The data that support the findings of this study are available from the corresponding author upon a reasonable request.
\end{suppinfo}


\nocite{*}


\begin{mcitethebibliography}{53}
\providecommand*\natexlab[1]{#1}
\providecommand*\mciteSetBstSublistMode[1]{}
\providecommand*\mciteSetBstMaxWidthForm[2]{}
\providecommand*\mciteBstWouldAddEndPuncttrue
  {\def\EndOfBibitem{\unskip.}}
\providecommand*\mciteBstWouldAddEndPunctfalse
  {\let\EndOfBibitem\relax}
\providecommand*\mciteSetBstMidEndSepPunct[3]{}
\providecommand*\mciteSetBstSublistLabelBeginEnd[3]{}
\providecommand*\EndOfBibitem{}
\mciteSetBstSublistMode{f}
\mciteSetBstMaxWidthForm{subitem}{(\alph{mcitesubitemcount})}
\mciteSetBstSublistLabelBeginEnd
  {\mcitemaxwidthsubitemform\space}
  {\relax}
  {\relax}

\bibitem[Briegel \latin{et~al.}(1998)Briegel, Dür, Cirac, and
  Zoller]{briegel_quantum_1998}
Briegel,~H.-J.; Dür,~W.; Cirac,~J.~I.; Zoller,~P. Quantum {Repeaters}: {The}
  {Role} of {Imperfect} {Local} {Operations} in {Quantum} {Communication}.
  \emph{Physical Review Letters} \textbf{1998}, \emph{81}, 5932--5935\relax
\mciteBstWouldAddEndPuncttrue
\mciteSetBstMidEndSepPunct{\mcitedefaultmidpunct}
{\mcitedefaultendpunct}{\mcitedefaultseppunct}\relax
\EndOfBibitem
\bibitem[Sangouard \latin{et~al.}(2007)Sangouard, Simon, Minář, Zbinden,
  de~Riedmatten, and Gisin]{sangouard_long-distance_2007}
Sangouard,~N.; Simon,~C.; Minář,~J.; Zbinden,~H.; de~Riedmatten,~H.;
  Gisin,~N. Long-distance entanglement distribution with single-photon sources.
  \emph{Physical Review A} \textbf{2007}, \emph{76}, 050301\relax
\mciteBstWouldAddEndPuncttrue
\mciteSetBstMidEndSepPunct{\mcitedefaultmidpunct}
{\mcitedefaultendpunct}{\mcitedefaultseppunct}\relax
\EndOfBibitem
\bibitem[Lu and Pan(2014)Lu, and Pan]{lu_push-button_2014}
Lu,~C.-Y.; Pan,~J.-W. Push-button photon entanglement. \emph{Nature Photonics}
  \textbf{2014}, \emph{8}, 174--176\relax
\mciteBstWouldAddEndPuncttrue
\mciteSetBstMidEndSepPunct{\mcitedefaultmidpunct}
{\mcitedefaultendpunct}{\mcitedefaultseppunct}\relax
\EndOfBibitem
\bibitem[Wang \latin{et~al.}(2019)Wang, Hu, Chung, Qin, Yang, Li, Liu, Zhong,
  He, Ding, Deng, Dai, Huo, Höfling, Lu, and Pan]{wang_-demand_2019}
Wang,~H. \latin{et~al.}  On-{Demand} {Semiconductor} {Source} of {Entangled}
  {Photons} {Which} {Simultaneously} {Has} {High} {Fidelity}, {Efficiency}, and
  {Indistinguishability}. \emph{Physical Review Letters} \textbf{2019},
  \emph{122}, 113602\relax
\mciteBstWouldAddEndPuncttrue
\mciteSetBstMidEndSepPunct{\mcitedefaultmidpunct}
{\mcitedefaultendpunct}{\mcitedefaultseppunct}\relax
\EndOfBibitem
\bibitem[Aharonovich \latin{et~al.}(2016)Aharonovich, Englund, and
  Toth]{aharonovich_solid-state_2016}
Aharonovich,~I.; Englund,~D.; Toth,~M. Solid-state single-photon emitters.
  \emph{Nature Photonics} \textbf{2016}, \emph{10}, 631--641\relax
\mciteBstWouldAddEndPuncttrue
\mciteSetBstMidEndSepPunct{\mcitedefaultmidpunct}
{\mcitedefaultendpunct}{\mcitedefaultseppunct}\relax
\EndOfBibitem
\bibitem[Heindel \latin{et~al.}(2023)Heindel, Kim, Gregersen, Rastelli, and
  Reitzenstein]{heindel_quantum_2023}
Heindel,~T.; Kim,~J.-H.; Gregersen,~N.; Rastelli,~A.; Reitzenstein,~S. Quantum
  dots for photonic quantum information technology. \emph{Advances in Optics
  and Photonics} \textbf{2023}, \emph{15}, 613\relax
\mciteBstWouldAddEndPuncttrue
\mciteSetBstMidEndSepPunct{\mcitedefaultmidpunct}
{\mcitedefaultendpunct}{\mcitedefaultseppunct}\relax
\EndOfBibitem
\bibitem[Rodt \latin{et~al.}(2020)Rodt, Reitzenstein, and
  Heindel]{rodt_deterministically_2020}
Rodt,~S.; Reitzenstein,~S.; Heindel,~T. Deterministically fabricated
  solid-state quantum-light sources. \emph{Journal of Physics: Condensed
  Matter} \textbf{2020}, \emph{32}, 153003\relax
\mciteBstWouldAddEndPuncttrue
\mciteSetBstMidEndSepPunct{\mcitedefaultmidpunct}
{\mcitedefaultendpunct}{\mcitedefaultseppunct}\relax
\EndOfBibitem
\bibitem[Patel \latin{et~al.}(2010)Patel, Bennett, Farrer, Nicoll, Ritchie, and
  Shields]{patel_two-photon_2010}
Patel,~R.~B.; Bennett,~A.~J.; Farrer,~I.; Nicoll,~C.~A.; Ritchie,~D.~A.;
  Shields,~A.~J. Two-photon interference of the emission from electrically
  tunable remote quantum dots. \emph{Nature Photonics} \textbf{2010}, \emph{4},
  632--635\relax
\mciteBstWouldAddEndPuncttrue
\mciteSetBstMidEndSepPunct{\mcitedefaultmidpunct}
{\mcitedefaultendpunct}{\mcitedefaultseppunct}\relax
\EndOfBibitem
\bibitem[Zhai \latin{et~al.}(2022)Zhai, Nguyen, Spinnler, Ritzmann, Löbl,
  Wieck, Ludwig, Javadi, and Warburton]{zhai_quantum_2022}
Zhai,~L.; Nguyen,~G.~N.; Spinnler,~C.; Ritzmann,~J.; Löbl,~M.~C.;
  Wieck,~A.~D.; Ludwig,~A.; Javadi,~A.; Warburton,~R.~J. Quantum {Interference}
  of {Identical} {Photons} from {Remote} {GaAs} {Quantum} {Dots}. \emph{Nature
  Nanotechnology} \textbf{2022}, \emph{17}, 829--833, arXiv:2106.03871
  [cond-mat, physics:quant-ph]\relax
\mciteBstWouldAddEndPuncttrue
\mciteSetBstMidEndSepPunct{\mcitedefaultmidpunct}
{\mcitedefaultendpunct}{\mcitedefaultseppunct}\relax
\EndOfBibitem
\bibitem[Warburton \latin{et~al.}(1997)Warburton, Dürr, Karrai, Kotthaus,
  Medeiros-Ribeiro, and Petroff]{warburton_charged_1997}
Warburton,~R.~J.; Dürr,~C.~S.; Karrai,~K.; Kotthaus,~J.~P.;
  Medeiros-Ribeiro,~G.; Petroff,~P.~M. Charged {Excitons} in {Self}-{Assembled}
  {Semiconductor} {Quantum} {Dots}. \emph{Physical Review Letters}
  \textbf{1997}, \emph{79}, 5282--5285\relax
\mciteBstWouldAddEndPuncttrue
\mciteSetBstMidEndSepPunct{\mcitedefaultmidpunct}
{\mcitedefaultendpunct}{\mcitedefaultseppunct}\relax
\EndOfBibitem
\bibitem[Hermannstädter \latin{et~al.}(2009)Hermannstädter, Witzany, Beirne,
  Schulz, Eichfelder, Rossbach, Jetter, Michler, Wang, Rastelli, and
  Schmidt]{hermannstadter_polarization_2009}
Hermannstädter,~C.; Witzany,~M.; Beirne,~G.~J.; Schulz,~W.-M.; Eichfelder,~M.;
  Rossbach,~R.; Jetter,~M.; Michler,~P.; Wang,~L.; Rastelli,~A.; Schmidt,~O.~G.
  Polarization fine structure and enhanced single-photon emission of
  self-assembled lateral {InGaAs} quantum dot molecules embedded in a planar
  microcavity. \emph{Journal of Applied Physics} \textbf{2009}, \emph{105},
  122408\relax
\mciteBstWouldAddEndPuncttrue
\mciteSetBstMidEndSepPunct{\mcitedefaultmidpunct}
{\mcitedefaultendpunct}{\mcitedefaultseppunct}\relax
\EndOfBibitem
\bibitem[Löbl \latin{et~al.}(2017)Löbl, Söllner, Javadi, Pregnolato, Schott,
  Midolo, Kuhlmann, Stobbe, Wieck, Lodahl, Ludwig, and
  Warburton]{lobl_narrow_2017}
Löbl,~M.~C.; Söllner,~I.; Javadi,~A.; Pregnolato,~T.; Schott,~R.; Midolo,~L.;
  Kuhlmann,~A.~V.; Stobbe,~S.; Wieck,~A.~D.; Lodahl,~P.; Ludwig,~A.;
  Warburton,~R.~J. Narrow optical linewidths and spin pumping on charge-tunable
  close-to-surface self-assembled quantum dots in an ultrathin diode.
  \emph{Physical Review B} \textbf{2017}, \emph{96}, 165440\relax
\mciteBstWouldAddEndPuncttrue
\mciteSetBstMidEndSepPunct{\mcitedefaultmidpunct}
{\mcitedefaultendpunct}{\mcitedefaultseppunct}\relax
\EndOfBibitem
\bibitem[Schall \latin{et~al.}(2021)Schall, Deconinck, Bart, Florian,
  Helversen, Dangel, Schmidt, Bremer, Bopp, Hüllen, Gies, Reuter, Wieck, Rodt,
  Finley, Jahnke, Ludwig, and Reitzenstein]{schall_bright_2021}
Schall,~J. \latin{et~al.}  Bright {Electrically} {Controllable}
  {Quantum}‐{Dot}‐{Molecule} {Devices} {Fabricated} by {In} {Situ}
  {Electron}‐{Beam} {Lithography}. \emph{Advanced Quantum Technologies}
  \textbf{2021}, \emph{4}, 2100002\relax
\mciteBstWouldAddEndPuncttrue
\mciteSetBstMidEndSepPunct{\mcitedefaultmidpunct}
{\mcitedefaultendpunct}{\mcitedefaultseppunct}\relax
\EndOfBibitem
\bibitem[Miller \latin{et~al.}(1984)Miller, Chemla, Damen, Gossard, Wiegmann,
  Wood, and Burrus]{miller_band-edge_1984}
Miller,~D. A.~B.; Chemla,~D.~S.; Damen,~T.~C.; Gossard,~A.~C.; Wiegmann,~W.;
  Wood,~T.~H.; Burrus,~C.~A. Band-{Edge} {Electroabsorption} in {Quantum}
  {Well} {Structures}: {The} {Quantum}-{Confined} {Stark} {Effect}.
  \emph{Physical Review Letters} \textbf{1984}, \emph{53}, 2173--2176\relax
\mciteBstWouldAddEndPuncttrue
\mciteSetBstMidEndSepPunct{\mcitedefaultmidpunct}
{\mcitedefaultendpunct}{\mcitedefaultseppunct}\relax
\EndOfBibitem
\bibitem[Bennett \latin{et~al.}(2010)Bennett, Patel, Skiba-Szymanska, Nicoll,
  Farrer, Ritchie, and Shields]{bennett_giant_2010}
Bennett,~A.~J.; Patel,~R.~B.; Skiba-Szymanska,~J.; Nicoll,~C.~A.; Farrer,~I.;
  Ritchie,~D.~A.; Shields,~A.~J. Giant {Stark} effect in the emission of single
  semiconductor quantum dots. \emph{Applied Physics Letters} \textbf{2010},
  \emph{97}, 031104\relax
\mciteBstWouldAddEndPuncttrue
\mciteSetBstMidEndSepPunct{\mcitedefaultmidpunct}
{\mcitedefaultendpunct}{\mcitedefaultseppunct}\relax
\EndOfBibitem
\bibitem[Nowak \latin{et~al.}(2014)Nowak, Portalupi, Giesz, Gazzano, Dal~Savio,
  Braun, Karrai, Arnold, Lanco, Sagnes, Lemaître, and
  Senellart]{nowak_deterministic_2014}
Nowak,~A.~K.; Portalupi,~S.~L.; Giesz,~V.; Gazzano,~O.; Dal~Savio,~C.;
  Braun,~P.-F.; Karrai,~K.; Arnold,~C.; Lanco,~L.; Sagnes,~I.; Lemaître,~A.;
  Senellart,~P. Deterministic and electrically tunable bright single-photon
  source. \emph{Nature Communications} \textbf{2014}, \emph{5}, 3240\relax
\mciteBstWouldAddEndPuncttrue
\mciteSetBstMidEndSepPunct{\mcitedefaultmidpunct}
{\mcitedefaultendpunct}{\mcitedefaultseppunct}\relax
\EndOfBibitem
\bibitem[Schnauber \latin{et~al.}(2021)Schnauber, Große, Kaganskiy, Ott,
  Anikin, Schmidt, Rodt, and Reitzenstein]{schnauber_spectral_2021}
Schnauber,~P.; Große,~J.; Kaganskiy,~A.; Ott,~M.; Anikin,~P.; Schmidt,~R.;
  Rodt,~S.; Reitzenstein,~S. Spectral control of deterministically fabricated
  quantum dot waveguide systems using the quantum confined {Stark} effect.
  \emph{APL Photonics} \textbf{2021}, \emph{6}, 050801\relax
\mciteBstWouldAddEndPuncttrue
\mciteSetBstMidEndSepPunct{\mcitedefaultmidpunct}
{\mcitedefaultendpunct}{\mcitedefaultseppunct}\relax
\EndOfBibitem
\bibitem[Davanço \latin{et~al.}(2011)Davanço, Rakher, Schuh, Badolato, and
  Srinivasan]{davanco_circular_2011}
Davanço,~M.; Rakher,~M.~T.; Schuh,~D.; Badolato,~A.; Srinivasan,~K. A circular
  dielectric grating for vertical extraction of single quantum dot emission.
  \emph{Applied Physics Letters} \textbf{2011}, \emph{99}, 041102\relax
\mciteBstWouldAddEndPuncttrue
\mciteSetBstMidEndSepPunct{\mcitedefaultmidpunct}
{\mcitedefaultendpunct}{\mcitedefaultseppunct}\relax
\EndOfBibitem
\bibitem[Liu \latin{et~al.}(2019)Liu, Su, Wei, Yao, Silva, Yu, Iles-Smith,
  Srinivasan, Rastelli, Li, and Wang]{liu_solid-state_2019}
Liu,~J.; Su,~R.; Wei,~Y.; Yao,~B.; Silva,~S. F. C.~D.; Yu,~Y.; Iles-Smith,~J.;
  Srinivasan,~K.; Rastelli,~A.; Li,~J.; Wang,~X. A solid-state source of
  strongly entangled photon pairs with high brightness and
  indistinguishability. \emph{Nature Nanotechnology} \textbf{2019}, \emph{14},
  586--593\relax
\mciteBstWouldAddEndPuncttrue
\mciteSetBstMidEndSepPunct{\mcitedefaultmidpunct}
{\mcitedefaultendpunct}{\mcitedefaultseppunct}\relax
\EndOfBibitem
\bibitem[Ates \latin{et~al.}(2012)Ates, Sapienza, Davanco, Badolato, and
  Srinivasan]{ates_bright_2012}
Ates,~S.; Sapienza,~L.; Davanco,~M.; Badolato,~A.; Srinivasan,~K. Bright
  {Single}-{Photon} {Emission} {From} a {Quantum} {Dot} in a {Circular} {Bragg}
  {Grating} {Microcavity}. \emph{IEEE Journal of Selected Topics in Quantum
  Electronics} \textbf{2012}, \emph{18}, 1711--1721\relax
\mciteBstWouldAddEndPuncttrue
\mciteSetBstMidEndSepPunct{\mcitedefaultmidpunct}
{\mcitedefaultendpunct}{\mcitedefaultseppunct}\relax
\EndOfBibitem
\bibitem[Yao \latin{et~al.}(2018)Yao, Su, Wei, Liu, Zhao, and
  Liu]{yao_design_2018}
Yao,~B.; Su,~R.; Wei,~Y.; Liu,~Z.; Zhao,~T.; Liu,~J. Design for {Hybrid}
  {Circular} {Bragg} {Gratings} for a {Highly} {Efficient} {Quantum}-{Dot}
  {Single}-{Photon} {Source}. \emph{Journal of the Korean Physical Society}
  \textbf{2018}, \emph{73}, 1502--1505\relax
\mciteBstWouldAddEndPuncttrue
\mciteSetBstMidEndSepPunct{\mcitedefaultmidpunct}
{\mcitedefaultendpunct}{\mcitedefaultseppunct}\relax
\EndOfBibitem
\bibitem[Rickert \latin{et~al.}(2023)Rickert, Betz, Plock, Burger, and
  Heindel]{rickert_high-performance_2023}
Rickert,~L.; Betz,~F.; Plock,~M.; Burger,~S.; Heindel,~T. High-performance
  designs for fiber-pigtailed quantum-light sources based on quantum dots in
  electrically-controlled circular {Bragg} gratings. \emph{Optics Express}
  \textbf{2023}, \emph{31}, 14750\relax
\mciteBstWouldAddEndPuncttrue
\mciteSetBstMidEndSepPunct{\mcitedefaultmidpunct}
{\mcitedefaultendpunct}{\mcitedefaultseppunct}\relax
\EndOfBibitem
\bibitem[Ji \latin{et~al.}(2021)Ji, Tajiri, Kiyama, Oiwa, and
  Iwamoto]{ji_design_2021}
Ji,~S.; Tajiri,~T.; Kiyama,~H.; Oiwa,~A.; Iwamoto,~S. Design of bull’s-eye
  optical cavity toward efficient quantum media conversion using gate-defined
  quantum dot. \emph{Japanese Journal of Applied Physics} \textbf{2021},
  \emph{60}, 102003\relax
\mciteBstWouldAddEndPuncttrue
\mciteSetBstMidEndSepPunct{\mcitedefaultmidpunct}
{\mcitedefaultendpunct}{\mcitedefaultseppunct}\relax
\EndOfBibitem
\bibitem[Barbiero \latin{et~al.}(2022)Barbiero, Huwer, Skiba-Szymanska,
  Müller, Stevenson, and Shields]{barbiero_design_2022}
Barbiero,~A.; Huwer,~J.; Skiba-Szymanska,~J.; Müller,~T.; Stevenson,~R.~M.;
  Shields,~A.~J. Design study for an efficient semiconductor quantum light
  source operating in the telecom {C}-band based on an electrically-driven
  circular {Bragg} grating. \emph{Optics Express} \textbf{2022}, \emph{30},
  10919\relax
\mciteBstWouldAddEndPuncttrue
\mciteSetBstMidEndSepPunct{\mcitedefaultmidpunct}
{\mcitedefaultendpunct}{\mcitedefaultseppunct}\relax
\EndOfBibitem
\bibitem[Buchinger \latin{et~al.}(2023)Buchinger, Betzold, Höfling, and
  Huber-Loyola]{buchinger_optical_2023}
Buchinger,~Q.; Betzold,~S.; Höfling,~S.; Huber-Loyola,~T. Optical properties
  of circular {Bragg} gratings with labyrinth geometry to enable electrical
  contacts. \emph{Applied Physics Letters} \textbf{2023}, \emph{122},
  111110\relax
\mciteBstWouldAddEndPuncttrue
\mciteSetBstMidEndSepPunct{\mcitedefaultmidpunct}
{\mcitedefaultendpunct}{\mcitedefaultseppunct}\relax
\EndOfBibitem
\bibitem[Shih \latin{et~al.}(2023)Shih, Rodt, and Reitzenstein]{Shih2023}
Shih,~C.-W.; Rodt,~S.; Reitzenstein,~S. Universal design method for bright
  quantum light sources based on circular Bragg grating cavities. \emph{Optics
  Express} \textbf{2023}, \emph{31}, 35552\relax
\mciteBstWouldAddEndPuncttrue
\mciteSetBstMidEndSepPunct{\mcitedefaultmidpunct}
{\mcitedefaultendpunct}{\mcitedefaultseppunct}\relax
\EndOfBibitem
\bibitem[Schnauber \latin{et~al.}(2019)Schnauber, Singh, Schall, Park, Song,
  Rodt, Srinivasan, Reitzenstein, and
  Davanco]{schnauber_indistinguishable_2019}
Schnauber,~P.; Singh,~A.; Schall,~J.; Park,~S.~I.; Song,~J.~D.; Rodt,~S.;
  Srinivasan,~K.; Reitzenstein,~S.; Davanco,~M. Indistinguishable {Photons}
  from {Deterministically} {Integrated} {Single} {Quantum} {Dots} in
  {Heterogeneous} {GaAs}/{Si} $_{\textrm{3}}$ {N} $_{\textrm{4}}$ {Quantum}
  {Photonic} {Circuits}. \emph{Nano Letters} \textbf{2019}, \emph{19},
  7164--7172\relax
\mciteBstWouldAddEndPuncttrue
\mciteSetBstMidEndSepPunct{\mcitedefaultmidpunct}
{\mcitedefaultendpunct}{\mcitedefaultseppunct}\relax
\EndOfBibitem
\bibitem[Hoehne \latin{et~al.}(2019)Hoehne, Schnauber, Rodt, Reitzenstein, and
  Burger]{hoehne_numerical_2019}
Hoehne,~T.; Schnauber,~P.; Rodt,~S.; Reitzenstein,~S.; Burger,~S. Numerical
  {Investigation} of {Light} {Emission} from {Quantum} {Dots} {Embedded} into
  {On}‐{Chip}, {Low}‐{Index}‐{Contrast} {Optical} {Waveguides}.
  \emph{physica status solidi (b)} \textbf{2019}, \emph{256}, 1800437\relax
\mciteBstWouldAddEndPuncttrue
\mciteSetBstMidEndSepPunct{\mcitedefaultmidpunct}
{\mcitedefaultendpunct}{\mcitedefaultseppunct}\relax
\EndOfBibitem
\bibitem[Cole \latin{et~al.}(1992)Cole, Salimian, Cooper, Lee, and
  Dutta]{cole_reactive_1992}
Cole,~M.~W.; Salimian,~S.; Cooper,~C.~B.; Lee,~H.~S.; Dutta,~M. Reactive ion
  etching of {GaAs} with {SiCl} $_{\textrm{4}}$ : {A} residual damage and
  electrical investigation. \emph{Scanning} \textbf{1992}, \emph{14},
  31--36\relax
\mciteBstWouldAddEndPuncttrue
\mciteSetBstMidEndSepPunct{\mcitedefaultmidpunct}
{\mcitedefaultendpunct}{\mcitedefaultseppunct}\relax
\EndOfBibitem
\bibitem[Burger \latin{et~al.}(2015)Burger, Zschiedrich, Pomplun, Herrmann, and
  Schmidt]{burger_hp-finite_2015}
Burger,~S.; Zschiedrich,~L.; Pomplun,~J.; Herrmann,~S.; Schmidt,~F. hp-finite
  element method for simulating light scattering from complex {3D} structures.
  2015; p 94240Z, arXiv:1503.06617 [physics]\relax
\mciteBstWouldAddEndPuncttrue
\mciteSetBstMidEndSepPunct{\mcitedefaultmidpunct}
{\mcitedefaultendpunct}{\mcitedefaultseppunct}\relax
\EndOfBibitem
\bibitem[Li \latin{et~al.}(2023)Li, Yang, Schall, Von~Helversen, Palekar, Liu,
  Roche, Rodt, Ni, Zhang, Niu, and Reitzenstein]{li_scalable_2023}
Li,~S.; Yang,~Y.; Schall,~J.; Von~Helversen,~M.; Palekar,~C.; Liu,~H.;
  Roche,~L.; Rodt,~S.; Ni,~H.; Zhang,~Y.; Niu,~Z.; Reitzenstein,~S. Scalable
  {Deterministic} {Integration} of {Two} {Quantum} {Dots} into an {On}-{Chip}
  {Quantum} {Circuit}. \emph{ACS Photonics} \textbf{2023}, \emph{10},
  2846--2853\relax
\mciteBstWouldAddEndPuncttrue
\mciteSetBstMidEndSepPunct{\mcitedefaultmidpunct}
{\mcitedefaultendpunct}{\mcitedefaultseppunct}\relax
\EndOfBibitem
\bibitem[Fry \latin{et~al.}(2000)Fry, Finley, Wilson, Lemaître, Mowbray,
  Skolnick, Hopkinson, Hill, and Clark]{fry_electric-field-dependent_2000}
Fry,~P.~W.; Finley,~J.~J.; Wilson,~L.~R.; Lemaître,~A.; Mowbray,~D.~J.;
  Skolnick,~M.~S.; Hopkinson,~M.; Hill,~G.; Clark,~J.~C.
  Electric-field-dependent carrier capture and escape in self-assembled
  {InAs}/{GaAs} quantum dots. \emph{Applied Physics Letters} \textbf{2000},
  \emph{77}, 4344--4346\relax
\mciteBstWouldAddEndPuncttrue
\mciteSetBstMidEndSepPunct{\mcitedefaultmidpunct}
{\mcitedefaultendpunct}{\mcitedefaultseppunct}\relax
\EndOfBibitem
\bibitem[Oulton \latin{et~al.}(2002)Oulton, Finley, Ashmore, Gregory, Mowbray,
  Skolnick, Steer, Liew, Migliorato, and Cullis]{oulton_manipulation_2002}
Oulton,~R.; Finley,~J.~J.; Ashmore,~A.~D.; Gregory,~I.~S.; Mowbray,~D.~J.;
  Skolnick,~M.~S.; Steer,~M.~J.; Liew,~S.-L.; Migliorato,~M.~A.; Cullis,~A.~J.
  Manipulation of the homogeneous linewidth of an individual {In}({Ga}){As}
  quantum dot. \emph{Physical Review B} \textbf{2002}, \emph{66}, 045313\relax
\mciteBstWouldAddEndPuncttrue
\mciteSetBstMidEndSepPunct{\mcitedefaultmidpunct}
{\mcitedefaultendpunct}{\mcitedefaultseppunct}\relax
\EndOfBibitem
\bibitem[Finley \latin{et~al.}(2001)Finley, Ashmore, Lemaître, Mowbray,
  Skolnick, Itskevich, Maksym, Hopkinson, and Krauss]{finley_charged_2001}
Finley,~J.~J.; Ashmore,~A.~D.; Lemaître,~A.; Mowbray,~D.~J.; Skolnick,~M.~S.;
  Itskevich,~I.~E.; Maksym,~P.~A.; Hopkinson,~M.; Krauss,~T.~F. Charged and
  neutral exciton complexes in individual self-assembled {In} ( {Ga} ) {As}
  quantum dots. \emph{Physical Review B} \textbf{2001}, \emph{63}, 073307\relax
\mciteBstWouldAddEndPuncttrue
\mciteSetBstMidEndSepPunct{\mcitedefaultmidpunct}
{\mcitedefaultendpunct}{\mcitedefaultseppunct}\relax
\EndOfBibitem
\bibitem[Akimov \latin{et~al.}(2005)Akimov, Kavokin, Hundt, and
  Henneberger]{akimov_electron-hole_2005}
Akimov,~I.~A.; Kavokin,~K.~V.; Hundt,~A.; Henneberger,~F. Electron-hole
  exchange interaction in a negatively charged quantum dot. \emph{Physical
  Review B} \textbf{2005}, \emph{71}, 075326\relax
\mciteBstWouldAddEndPuncttrue
\mciteSetBstMidEndSepPunct{\mcitedefaultmidpunct}
{\mcitedefaultendpunct}{\mcitedefaultseppunct}\relax
\EndOfBibitem
\bibitem[Ediger \latin{et~al.}(2007)Ediger, Bester, Gerardot, Badolato,
  Petroff, Karrai, Zunger, and Warburton]{ediger_fine_2007}
Ediger,~M.; Bester,~G.; Gerardot,~B.~D.; Badolato,~A.; Petroff,~P.~M.;
  Karrai,~K.; Zunger,~A.; Warburton,~R.~J. Fine {Structure} of {Negatively} and
  {Positively} {Charged} {Excitons} in {Semiconductor} {Quantum} {Dots}:
  {Electron}-{Hole} {Asymmetry}. \emph{Physical Review Letters} \textbf{2007},
  \emph{98}, 036808\relax
\mciteBstWouldAddEndPuncttrue
\mciteSetBstMidEndSepPunct{\mcitedefaultmidpunct}
{\mcitedefaultendpunct}{\mcitedefaultseppunct}\relax
\EndOfBibitem
\bibitem[Warming \latin{et~al.}(2009)Warming, Siebert, Schliwa, Stock,
  Zimmermann, and Bimberg]{warming_hole-hole_2009}
Warming,~T.; Siebert,~E.; Schliwa,~A.; Stock,~E.; Zimmermann,~R.; Bimberg,~D.
  Hole-hole and electron-hole exchange interactions in single {InAs}/{GaAs}
  quantum dots. \emph{Physical Review B} \textbf{2009}, \emph{79}, 125316\relax
\mciteBstWouldAddEndPuncttrue
\mciteSetBstMidEndSepPunct{\mcitedefaultmidpunct}
{\mcitedefaultendpunct}{\mcitedefaultseppunct}\relax
\EndOfBibitem
\bibitem[Seguin \latin{et~al.}(2005)Seguin, Schliwa, Rodt, Pötschke, Pohl, and
  Bimberg]{seguin_size-dependent_2005}
Seguin,~R.; Schliwa,~A.; Rodt,~S.; Pötschke,~K.; Pohl,~U.~W.; Bimberg,~D.
  Size-{Dependent} {Fine}-{Structure} {Splitting} in {Self}-{Organized} {InAs}
  / {GaAs} {Quantum} {Dots}. \emph{Physical Review Letters} \textbf{2005},
  \emph{95}, 257402\relax
\mciteBstWouldAddEndPuncttrue
\mciteSetBstMidEndSepPunct{\mcitedefaultmidpunct}
{\mcitedefaultendpunct}{\mcitedefaultseppunct}\relax
\EndOfBibitem
\bibitem[Kavokin(2003)]{kavokin_fine_2003}
Kavokin,~K.~V. Fine structure of the quantum-dot trion. \emph{physica status
  solidi (a)} \textbf{2003}, \emph{195}, 592--595\relax
\mciteBstWouldAddEndPuncttrue
\mciteSetBstMidEndSepPunct{\mcitedefaultmidpunct}
{\mcitedefaultendpunct}{\mcitedefaultseppunct}\relax
\EndOfBibitem
\bibitem[Urbaszek \latin{et~al.}(2003)Urbaszek, Warburton, Karrai, Gerardot,
  Petroff, and Garcia]{urbaszek_fine_2003}
Urbaszek,~B.; Warburton,~R.~J.; Karrai,~K.; Gerardot,~B.~D.; Petroff,~P.~M.;
  Garcia,~J.~M. Fine {Structure} of {Highly} {Charged} {Excitons} in
  {Semiconductor} {Quantum} {Dots}. \emph{Physical Review Letters}
  \textbf{2003}, \emph{90}, 247403\relax
\mciteBstWouldAddEndPuncttrue
\mciteSetBstMidEndSepPunct{\mcitedefaultmidpunct}
{\mcitedefaultendpunct}{\mcitedefaultseppunct}\relax
\EndOfBibitem
\bibitem[Baier \latin{et~al.}(2001)Baier, Findeis, Zrenner, Bichler, and
  Abstreiter]{baier_optical_2001}
Baier,~M.; Findeis,~F.; Zrenner,~A.; Bichler,~M.; Abstreiter,~G. Optical
  spectroscopy of charged excitons in single quantum dot photodiodes.
  \emph{Physical Review B} \textbf{2001}, \emph{64}, 195326\relax
\mciteBstWouldAddEndPuncttrue
\mciteSetBstMidEndSepPunct{\mcitedefaultmidpunct}
{\mcitedefaultendpunct}{\mcitedefaultseppunct}\relax
\EndOfBibitem
\bibitem[Warburton \latin{et~al.}(2000)Warburton, Schäflein, Haft, Bickel,
  Lorke, Karrai, Garcia, Schoenfeld, and Petroff]{warburton_optical_2000}
Warburton,~R.~J.; Schäflein,~C.; Haft,~D.; Bickel,~F.; Lorke,~A.; Karrai,~K.;
  Garcia,~J.~M.; Schoenfeld,~W.; Petroff,~P.~M. Optical emission from a
  charge-tunable quantum ring. \emph{Nature} \textbf{2000}, \emph{405},
  926--929\relax
\mciteBstWouldAddEndPuncttrue
\mciteSetBstMidEndSepPunct{\mcitedefaultmidpunct}
{\mcitedefaultendpunct}{\mcitedefaultseppunct}\relax
\EndOfBibitem
\bibitem[Luo \latin{et~al.}(2019)Luo, Sun, Karasahin, Bracker, Carter, Yakes,
  Gammon, and Waks]{luo_spinphoton_2019}
Luo,~Z.; Sun,~S.; Karasahin,~A.; Bracker,~A.~S.; Carter,~S.~G.; Yakes,~M.~K.;
  Gammon,~D.; Waks,~E. A {Spin}–{Photon} {Interface} {Using}
  {Charge}-{Tunable} {Quantum} {Dots} {Strongly} {Coupled} to a {Cavity}.
  \emph{Nano Letters} \textbf{2019}, \emph{19}, 7072--7077\relax
\mciteBstWouldAddEndPuncttrue
\mciteSetBstMidEndSepPunct{\mcitedefaultmidpunct}
{\mcitedefaultendpunct}{\mcitedefaultseppunct}\relax
\EndOfBibitem
\bibitem[Neuwirth \latin{et~al.}(2021)Neuwirth, Basso~Basset, Rota, Roccia,
  Schimpf, Jöns, Rastelli, and Trotta]{neuwirth_quantum_2021}
Neuwirth,~J.; Basso~Basset,~F.; Rota,~M.~B.; Roccia,~E.; Schimpf,~C.;
  Jöns,~K.~D.; Rastelli,~A.; Trotta,~R. Quantum dot technology for quantum
  repeaters: from entangled photon generation toward the integration with
  quantum memories. \emph{Materials for Quantum Technology} \textbf{2021},
  \emph{1}, 043001\relax
\mciteBstWouldAddEndPuncttrue
\mciteSetBstMidEndSepPunct{\mcitedefaultmidpunct}
{\mcitedefaultendpunct}{\mcitedefaultseppunct}\relax
\EndOfBibitem
\bibitem[Azuma \latin{et~al.}(2015)Azuma, Tamaki, and
  Lo]{azuma_all-photonic_2015}
Azuma,~K.; Tamaki,~K.; Lo,~H.-K. All-photonic quantum repeaters. \emph{Nature
  Communications} \textbf{2015}, \emph{6}, 6787\relax
\mciteBstWouldAddEndPuncttrue
\mciteSetBstMidEndSepPunct{\mcitedefaultmidpunct}
{\mcitedefaultendpunct}{\mcitedefaultseppunct}\relax
\EndOfBibitem
\bibitem[Bopp \latin{et~al.}(2022)Bopp, Rojas, Revenga, Riedl, Sbresny, Boos,
  Simmet, Ahmadi, Gershoni, Kasprzak, Ludwig, Reitzenstein, Wieck, Reuter,
  Müller, and Finley]{bopp_quantum_2022}
Bopp,~F. \latin{et~al.}  Quantum {Dot} {Molecule} {Devices} with {Optical}
  {Control} of {Charge} {Status} and {Electronic} {Control} of {Coupling}.
  \emph{Advanced Quantum Technologies} \textbf{2022}, \emph{5}, 2200049\relax
\mciteBstWouldAddEndPuncttrue
\mciteSetBstMidEndSepPunct{\mcitedefaultmidpunct}
{\mcitedefaultendpunct}{\mcitedefaultseppunct}\relax
\EndOfBibitem
\bibitem[Vezvaee \latin{et~al.}(2022)Vezvaee, Hilaire, Doty, and
  Economou]{vezvaee_deterministic_2022}
Vezvaee,~A.; Hilaire,~P.; Doty,~M.~F.; Economou,~S.~E. Deterministic generation
  of entangled photonic cluster states from quantum dot molecules. 2022;
  \url{http://arxiv.org/abs/2206.03647}, arXiv:2206.03647 [quant-ph]\relax
\mciteBstWouldAddEndPuncttrue
\mciteSetBstMidEndSepPunct{\mcitedefaultmidpunct}
{\mcitedefaultendpunct}{\mcitedefaultseppunct}\relax
\EndOfBibitem
\bibitem[Madigawa \latin{et~al.}(2024)Madigawa, Donges, Gaál, Li, Jacobsen,
  Liu, Dai, Su, Shang, Ni, Schall, Rodt, Niu, Gregersen, Reitzenstein, and
  Munkhbat]{madigawa_assessing_2024}
Madigawa,~A.~A. \latin{et~al.}  Assessing the {Alignment} {Accuracy} of
  {State}-of-the-{Art} {Deterministic} {Fabrication} {Methods} for {Single}
  {Quantum} {Dot} {Devices}. \emph{ACS Photonics} \textbf{2024}, \emph{11},
  1012--1023\relax
\mciteBstWouldAddEndPuncttrue
\mciteSetBstMidEndSepPunct{\mcitedefaultmidpunct}
{\mcitedefaultendpunct}{\mcitedefaultseppunct}\relax
\EndOfBibitem
\bibitem[Hanbury~Brown and Twiss(1956)Hanbury~Brown, and
  Twiss]{hanbury_brown_question_1956}
Hanbury~Brown,~R.; Twiss,~R.~Q. The {Question} of {Correlation} between
  {Photons} in {Coherent} {Light} {Rays}. \emph{Nature} \textbf{1956},
  \emph{178}, 1447--1448\relax
\mciteBstWouldAddEndPuncttrue
\mciteSetBstMidEndSepPunct{\mcitedefaultmidpunct}
{\mcitedefaultendpunct}{\mcitedefaultseppunct}\relax
\EndOfBibitem
\bibitem[Ding \latin{et~al.}(2016)Ding, He, Duan, Gregersen, Chen, Unsleber,
  Maier, Schneider, Kamp, Höfling, Lu, and Pan]{ding_-demand_2016}
Ding,~X.; He,~Y.; Duan,~Z.-C.; Gregersen,~N.; Chen,~M.-C.; Unsleber,~S.;
  Maier,~S.; Schneider,~C.; Kamp,~M.; Höfling,~S.; Lu,~C.-Y.; Pan,~J.-W.
  On-{Demand} {Single} {Photons} with {High} {Extraction} {Efficiency} and
  {Near}-{Unity} {Indistinguishability} from a {Resonantly} {Driven} {Quantum}
  {Dot} in a {Micropillar}. \emph{Physical Review Letters} \textbf{2016},
  \emph{116}, 020401\relax
\mciteBstWouldAddEndPuncttrue
\mciteSetBstMidEndSepPunct{\mcitedefaultmidpunct}
{\mcitedefaultendpunct}{\mcitedefaultseppunct}\relax
\EndOfBibitem
\bibitem[Hong \latin{et~al.}(1987)Hong, Ou, and Mandel]{hong_measurement_1987}
Hong,~C.~K.; Ou,~Z.~Y.; Mandel,~L. Measurement of subpicosecond time intervals
  between two photons by interference. \emph{Physical Review Letters}
  \textbf{1987}, \emph{59}, 2044--2046\relax
\mciteBstWouldAddEndPuncttrue
\mciteSetBstMidEndSepPunct{\mcitedefaultmidpunct}
{\mcitedefaultendpunct}{\mcitedefaultseppunct}\relax
\EndOfBibitem
\bibitem[Manna \latin{et~al.}(2020)Manna, Huang, Da~Silva, Schimpf, Rota,
  Lehner, Reindl, Trotta, and Rastelli]{manna_surface_2020}
Manna,~S.; Huang,~H.; Da~Silva,~S. F.~C.; Schimpf,~C.; Rota,~M.~B.; Lehner,~B.;
  Reindl,~M.; Trotta,~R.; Rastelli,~A. Surface passivation and oxide
  encapsulation to improve optical properties of a single {GaAs} quantum dot
  close to the surface. \emph{Applied Surface Science} \textbf{2020},
  \emph{532}, 147360\relax
\mciteBstWouldAddEndPuncttrue
\mciteSetBstMidEndSepPunct{\mcitedefaultmidpunct}
{\mcitedefaultendpunct}{\mcitedefaultseppunct}\relax
\EndOfBibitem
\end{mcitethebibliography}

\providecommand{\latin}[1]{#1}
\makeatletter
\providecommand{\doi}
  {\begingroup\let\do\@makeother\dospecials
  \catcode`\{=1 \catcode`\}=2 \doi@aux}
\providecommand{\doi@aux}[1]{\endgroup\texttt{#1}}
\makeatother
\providecommand*\mcitethebibliography{\thebibliography}
\csname @ifundefined\endcsname{endmcitethebibliography}
  {\let\endmcitethebibliography\endthebibliography}{}

\end{document}